\pdfoutput=1
\documentclass[aip,jcp,reprint,floatfix]{revtex4-2}
\usepackage{amsmath,amssymb}
\usepackage{graphicx}
\usepackage{dcolumn}
\usepackage{bm}
\usepackage{multirow}
\usepackage{float}
\usepackage{hhline}
\usepackage{subfigure}
\usepackage{afterpage}
\usepackage{tabularx}
\usepackage{comment}
\usepackage{float}

\usepackage[utf8]{inputenc}
\usepackage[T1]{fontenc}
\usepackage{mathptmx}
\usepackage{physics}

\newcommand{\tensorg}{\overline{\overline{\mathbf{G}}}}

\newcommand\numberthis{\addtocounter{equation}{1}\tag{\theequation}}

\usepackage[dvipsnames]{xcolor}

\graphicspath{{Figures/}}

\begin{document}

\preprint{AIP/123-QED}

\title{Theory of Molecular Emission Power Spectra. III. Non-Hermitian Interactions in Multichromophoric Systems Coupled with Polaritons}

\author{Siwei Wang}
\affiliation{Institute of Atomic and Molecular Sciences, Academia Sinica, Taipei 10617, Taiwan}

\author{Jia-Liang Huang}
\affiliation{Institute of Atomic and Molecular Sciences, Academia Sinica, Taipei 10617, Taiwan}
\affiliation{Department of Chemistry, National Taiwan University, Taipei 10617, Taiwan}
\author{Liang-Yan Hsu}
\email{lyhsu@gate.sinica.edu.tw}
\affiliation{Institute of Atomic and Molecular Sciences, Academia Sinica, Taipei 10617, Taiwan}
\affiliation{Department of Chemistry, National Taiwan University, Taipei 10617, Taiwan}
\affiliation{National Center for Theoretical Sciences, Taipei 10617, Taiwan}

\begin{abstract}
Based on our previous study [S. Wang \textit{et al}. J. Chem. Phys. \textbf{153}, 184102 (2020)], we generalize the theory of molecular emission power spectra (EPS) from one molecule to multichromophoric systems in the framework of macroscopic quantum electrodynamics. This generalized theory is applicable to ensembles of molecules, providing a comprehensive description of the molecular spontaneous emission spectrum in arbitrary inhomogeneous, dispersive, and absorbing media. In the far-field region, the analytical formula of EPS can be expressed as the product of a lineshape function (LF) and an electromagnetic environment factor (EEF). 
To demonstrate the polaritonic effect on multichromophoric systems, we simulate the LF and EEF for one to three molecules weakly coupled to surface plasmon polaritons above a silver surface. Our analytical expressions show that the peak broadening originates from not only the spontaneous emission rates but also the imaginary part of resonant dipole-dipole interactions (non-Hermitian interactions), which is associated with the superradiance of molecular aggregates, indicating that the superradiance rate can be controlled through an intermolecular distance and the design of dielectric environments. This study presents an alternative
approach to directly analyze the hybrid-state dynamics of multichromophoric systems coupled with polaritons.
\end{abstract}


\maketitle

\section{Introduction}
\label{Sec:Introduction}
The coupling between photons and molecules has received extensive attention in physical chemistry and chemical physics over the past decades due to various promising experimental breakthroughs\cite{Ellenbogen2011,Hutchison2012,Vasa2013,Chikkaraddy2016,Stranius2018,thomas2019tilting,Xiang2019,Xiang2020}. Experimentally, light-matter hybrid states can be achieved by modifying the photonic environments using Fabry-P\'{e}rot cavities\cite{Hutchison2012,thomas2019tilting,Xiang2019,Xiang2020}, waveguides\cite{Ellenbogen2011,zeng2016,abir_2020}, photonic crystals\cite{Dintinger2005,wu_2018}, and plasmonic materials\cite{kasani_2019,Mueller2020}. Several pioneering experiments have demonstrated that both vibrational and electronic strong coupling with quantum light can be used to manipulate the optical properties and reaction pathways of molecules,  known as polaritonic chemistry\cite{Galego2017,Martinez2018,Ribeiro2018,Hertzog2019,Mandal2019,Nagarajan_2021,ebbesen_2023,bhuyan_2023,Mandal2023}, quantum electrodynamic chemistry\cite{Flick2017,Schafer2019,Wei2022,ruggenthaler_2023,wei_2024}, or cavity chemistry\cite{Herrera2016,A_Rubio2023}. Among these experiments, spectroscopic techniques have played a crucial role in providing clear evidence for the formation of these hybrid states. In most experiments, transmission spectra\cite{thomas2019tilting,Xiang2020} and dark-field scattering spectra\cite{Chikkaraddy2016,Timur_2019,Heintz_2021,gupta_2021} are used to identify whether the hybrid states are formed. In addition to these experimental techniques, from a theoretical point of view, one alternative approach is to directly analyze the emission power spectra of molecules, which may offer insights into the hybrid-state dynamics.

To better understand the emission power spectra of molecules in complex dielectric environments, we have developed relevant theories for a single molecule\cite{wang2020theory,lee_2021} based on macroscopic quantum electrodynamics (MQED)\cite{gruner1996,Dung1998,scheel_1999,Dung2000,Dung2002_Energy}. According to our theory, the emission power spectrum of a single molecule consists of the lineshape function (LF) and the electromagnetic environment factor (EEF). The LF characterizes the quantum dynamics of hybrid states, while the EEF accounts for the spectral signatures contributed by dielectric environments. Our previous work demonstrated the interplay between exciton-photon and exciton-phonon (vibration) interactions\cite{wang2020theory}, as well as the impact of electromagnetic environments on spectra\cite{lee_2021}. However, our previous work is limited to one molecule case and does not address the case of multiple molecules, which are regarded as a key factor for the formation of molecular polaritons. To address this issue, we develop a theory for the emission power spectra of the spontaneous emission process in a collection of molecules and numerically demonstrate the emission power spectra of molecular emitters coupled with plasmon polaritons above a silver surface. 

Our article is organized as follows. In Sec.~\ref{Sec_Method_Hamil}, we briefly introduce the macroscopic QED Hamiltonian used in our previous works\cite{chuang_2024,chuang_2024_PRL}, and this Hamiltonian allows us to describe the light-matter interactions of multiple
emitters in complex dielectric environments. In Sec.~\ref{Sec_State_Vector}, we revisit our previous work on the wavefunction ansatz and the quantum dynamics of multichromophoric systems\cite{chuang_2024}, which serves as the foundation for evaluating LFs. In Sec.~\ref{Sec_Formalism_EPS}, we develop the theory for the emission power spectra of the spontaneous emission process in multichromophoric
systems and the key result can be expressed as a sum of the products of analytical EEFs and LFs. In Sec.~\ref{Sec_Approx_EPS}, we derive the approximate expression of LFs by evaluating the analytical LFs based on the wavefunction ansatz. Additionally, employing the Kramers–Kronig relations, we simplify the analytical EEFs and obtain the approximate EEFs. In Sec.~\ref{Sec:EPS_Planar}, we further simplify the approximate EEFs and LFs and obtain the specific EPS for planar systems. In Sec.~\ref{Sec:Discussion}, in order to explore the polaritonic effect on multichromophoric systems, we numerically demonstrate the LFs and EEFs from one to three molecules above a silver surface in the regime of weak light-matter interactions. In Sec.~\ref{Sec:Conclusion}, we summarize the main results of the current work and provide future perspectives.


\section{Method}
\label{Sec:Method}
\subsection{Hamiltonian}
\label{Sec_Method_Hamil}

\begin{figure}[htbp] \includegraphics[width=0.48\textwidth]{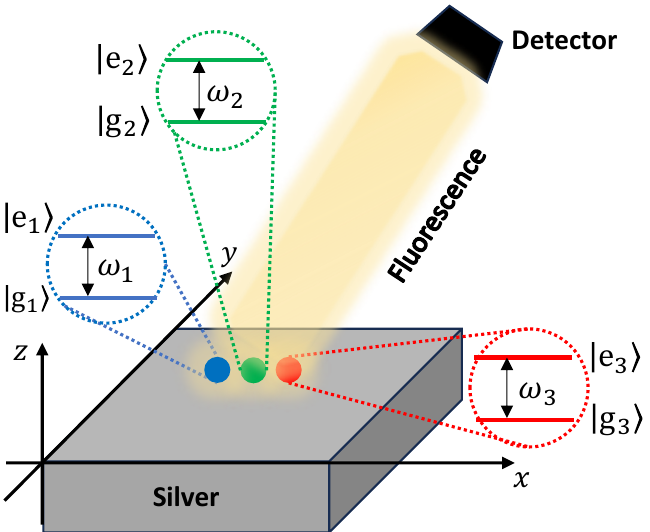}
\caption{Schematic illustration of the setup for measuring the emission power spectra of multiple emitters. Three molecules (represented by red, green, and blue dots) with electronic transition frequencies $\omega_1$, $\omega_2$, and $\omega_3$ are positioned above a silver surface. The emitted photons from the molecular aggregate are detected by a detector (represented by a black trapezoid) typically located in the far-field region. }
\label{Fig_1}
\end{figure}

In this study, we investigate the spontaneous emission of a molecular aggregate coupled to polaritons (dressed photons) at low temperatures. The setup for measuring the emission power spectra is shown in Fig.~\ref{Fig_1}, where multiple two-level molecules interact with polaritons in an arbitrary inhomogeneous, dispersive, and absorbing medium. To focus on the influence of the dielectric environment, we neglect phonon interactions (negligible at low temperatures)\cite{chuang_2024_PRL,spano_1989}, charge transfer effects (negligible at sufficient intermolecular distances)\cite{hestand_2017}, and disorder\cite{chen_2022}. Note that the simplification may limit the model to capture some features such as vibrational substructures and broad spectral bandwidths observed in real molecules. According to MQED in the multipolar coupling framework\cite{scully_1997,buhmann2013dispersion_1,snchezbarquilla_2022}, the total Hamiltonian $\hat{H}_\mathrm{T}$ (without the rotating wave approximation) under the electric dipole approximation\cite{chuang_2024,chuang_2024_2} can be expressed as $ \hat{H}_\mathrm{T} = \hat{H}_\mathrm{M} + \hat{H}_\mathrm{P} + \hat{H}_\mathrm{M-P} $, where the molecular Hamiltonian $\hat{H}_\mathrm{M}$, the polaritonic Hamiltonian $\hat{H}_\mathrm{P}$, and their interaction Hamiltonian $\hat{H}_\mathrm{M-P} $ are expressed as follows:
\begin{align}
\label{Eq:MQED_Molecules}
    &\hat{H}_\mathrm{M} = \sum_{\alpha=1}^{N_\mathrm{M}}  \hbar \omega_\alpha \hat{\sigma}_\alpha^{(+)} \hat{\sigma}_\alpha^{(-)}, \\   
\label{Eq:MQED_Polariton}    
    &\hat{H}_\mathrm{P} =  \int \mathrm{d}\mathbf{r}   \int_0^\infty d\omega \, \hbar \omega \, \hat{\mathbf{f}}^\dagger(\mathbf{r},\omega) \cdot \hat{\mathbf{f}}(\mathbf{r},\omega), \\
    & \hat{H}_\mathrm{M-P} = -\sum_{\alpha=1}^{N_\mathrm{M}} \hat{\pmb{\mu}}_\alpha \cdot \hat{\mathbf{E}}(\mathbf{r}_\alpha).
\label{Eq:Light-Matter_Interaction}
\end{align}
Here, $N_\mathrm{M}$ and $\alpha$ stand for the total number of molecules and the index of the molecules, respectively. In Eq.~(\ref{Eq:MQED_Molecules}), $\omega_\alpha$, $\hat{\sigma}_\alpha^{(+)}$ and $\hat{\sigma}_\alpha^{(-)}$ denote the electronic transition frequency, the raising operator, and the lowering operator of the $\alpha$-th molecule, respectively. In Eq.~(\ref{Eq:MQED_Polariton}), $\hat{\mathbf{f}}(\mathbf{r},\omega)$ and $\hat{\mathbf{f}}^\dagger(\mathbf{r},\omega) $ are the annihilation and creation operators of the bosonic vector fields\cite{Dung1998,Dung2002_Energy}. Moreover, their vector components satisfy the commutation relations: $[\hat{f}_i(\mathbf{r},\omega),\hat{f}_{j}^\dagger(\mathbf{r'},\omega')] = \delta_{i,j} \delta(\mathbf{r-r'}) \delta(\omega-\omega')$. In Eq.~(\ref{Eq:Light-Matter_Interaction}), $\hat{\pmb{\mu}}_\alpha$ and $\hat{\mathbf{E}}(\mathbf{r}_\alpha)$ are the transition dipole operator of the $\alpha$-th molecule and
the field operator at the molecular position $\mathbf{r}_\alpha$, respectively. The transition dipole operator $\hat{\pmb{\mu}}_\alpha$ can be expressed in terms of ${\pmb{\mu}}_\alpha $ and $\hat{\bar{\mu}}_\alpha $, i.e.,
    $\hat{\pmb{\mu}}_\alpha = {\pmb{\mu}}_\alpha \hat{\bar{\mu}}_\alpha$,  
where $\hat{\bar{\mu}}_\alpha =\hat{\sigma}_\alpha^{(+)} + \hat{\sigma}_\alpha^{(-)} $, and ${\pmb{\mu}}_\alpha$ is the electronic transition dipole moment of the $\alpha$-th molecule.
The field operator is expressed as: 
\begin{align}
    \hat{\mathbf{E}}(\mathbf{r}_\alpha) = \hat{\mathbf{E}}^{(+)}(\mathbf{r}_\alpha) + \hat{\mathbf{E}}^{(-)}(\mathbf{r}_\alpha),
\end{align}
with
\begin{align}
\label{Eq:E(+)}
    &\hat{\mathbf{E}}^{(+)}(\mathbf{r}_\alpha) =     \int_0^\infty d\omega \int \mathrm{d}\mathbf{r} \;  \overline{\overline{\mathbf{g}}}(\mathbf{r}_\alpha,\mathbf{r},\omega) \cdot \hat{\mathbf{f}}(\mathbf{r},\omega),   \\ 
\label{Eq:E(-)}
    &\hat{\mathbf{E}}^{(-)}(\mathbf{r}_\alpha) = \left\{\hat{\mathbf{E}}^{(+)}(\mathbf{r}_\alpha) \right\}^\dagger, \\
    &\overline{\overline{\mathbf{g}}}(\mathbf{r}_\alpha,\mathbf{r},\omega) = i\sqrt{ \frac{\hbar}{\pi\epsilon_0} }  \frac{\omega^2}{c^2} \sqrt{\mathrm{Im}\left[ \epsilon_\mathrm{r}(\mathbf{r},\omega) \right] } \, \overline{\overline{\mathbf{G}}}(\mathbf{r}_\alpha,\mathbf{r},\omega),
\end{align}
where $\epsilon_0$, $\epsilon_\mathrm{r}(\mathbf{r},\omega)$, and $c$ are the vacuum permittivity, the relative permittivity, and the speed of light in vacuum, respectively. In Eq.~(\ref{Eq:E(+)}), $  \overline{\overline{\mathbf{g}}}(\mathbf{r}_\alpha,\mathbf{r},\omega)$ is an auxiliary tensor function related to the dyadic Green's function $\overline{\overline{\mathbf{G}}}(\mathbf{r}_\alpha,\mathbf{r},\omega)$ that satisfies the macroscopic Maxwell's equation\cite{chew1995waves,novotny2012principles}:
\begin{align}
    \left[ \frac{\omega^2}{c^2}\varepsilon_\mathrm{r}(\mathbf{r}_\alpha,\omega) - \nabla \times \nabla \times \right] \overline{\overline{\mathbf{G}}}(\mathbf{r}_\alpha,\mathbf{r},\omega) = -\mathbf{\overline{\overline{I}}}_3 \delta(\mathbf{r}_\alpha-\mathbf{r}),
\label{Eq:GreensFunction_Maxwell}
\end{align}
where $\mathbf{\overline{\overline{I}}}_3$ is the $3 \times 3$ identity tensor, and $\delta(\mathbf{r}_\alpha-\mathbf{r})$ is the three-dimensional delta function.

\subsection{Truncated State Vector and Quantum Dynamics} 
\label{Sec_State_Vector}

The quantum dynamics of excited molecules can be extracted from the emission power spectrum, as the dynamical information is embedded in the lineshape functions, which will be discussed later. Therefore, to study the dynamics of molecular emitters, we begin with the time-dependent Schr\"{o}dinger equation $i\hbar\partial{\ket{\Psi(t)}}/\partial{t}=\hat{H}_\mathrm{T}\ket{\Psi(t)}$ and adopt a truncated state vector (i.e., a truncated wavefunction ansatz) based on our previous study\cite{chuang_2024}. The use of the truncated state vector includes the effects of counter-rotating interactions and captures the quantum dynamics for molecules coupled with polaritons above a silver surface in the weak and strong light-matter coupling regimes\cite{chuang_2024,chuang_2024_2}. The truncated state vector can be expressed as follows:
\begin{widetext}
\begin{align}
\nonumber
    \ket{\Psi(t)} =& \, C^{\mathrm{G},\{0\}}(t)\ket{\mathrm{G}} \ket{\left\{0\right\}}  + \sum_{\alpha=1}^{N_\mathrm{M}} C^{\mathrm{E}_\alpha,\{0\}}(t) e^{-i W^{\mathrm{E}_\alpha,\{0\}} t}\ket{\mathrm{E}_\alpha} \ket{\left\{0\right\}} + \sum_{k=1}^3 \int {d} \mathbf{r} \int_0^{\infty} {d} \omega C^{\mathrm{G},\left\{1_k\right\}}(\mathbf{r}, \omega, t) e^{-i W^{\mathrm{G},\{1\}}(\omega) t} \ket{\mathrm{G}}\ket{\left\{1_k(\mathbf{r}, \omega)\right\}} \\
    & +\sum_{\alpha=1}^{N_\mathrm{M}} \sum_{\beta>\alpha}^{N_\mathrm{M}} \sum_{k=1}^3 \int d \mathbf{r} \int_0^{\infty} d\omega \, C^{\mathrm{E}_{\alpha \beta},\left\{1_k\right\}}(\mathbf{r}, \omega, t) e^{-i W^{\mathrm{E}_{\alpha \beta},\{1\}}(\omega) t}\ket{ \mathrm{E}_{\alpha \beta}} \ket{\left\{1_k(\mathbf{r}, \omega)\right\} },
\label{Eq:State_Vector}
\end{align}
\end{widetext}
with
\begin{subequations}
\begin{align}
&W^{\mathrm{E}_\alpha,\{0\}}=\omega_\alpha, \\
&W^{\mathrm{G},\{1\}}(\omega)=\omega, \\
&W^{\mathrm{E}_{\alpha \beta},\{1\}}(\omega)=\omega+\omega_\alpha+\omega_\beta .
\end{align}
\end{subequations}

The state vector in Eq.~(\ref{Eq:State_Vector}) includes both the molecular electronic and photonic degrees of freedom in the entire system. For the molecular part, the ket state $\ket{\mathrm{G}}$ indicates that all molecules are in their electronically ground states, i.e., $\ket{\mathrm{G}} = \ket{\mathrm{g}_1}\ket{\mathrm{g}_2}\dots\ket{\mathrm{g}_{N_\mathrm{M}}}$; the ket state $\ket{\mathrm{E}_\alpha}$ denotes that the $\alpha$-th molecule is in its electronically excited state, while the other molecules remain in their electronically ground states, i.e., $\ket{\mathrm{E}_\alpha} = \hat{\sigma}^{(+)}_\alpha \ket{\mathrm{G}} $; the ket
state $\ket{\mathrm{E}_{\alpha\beta}} $ denotes that both the $\alpha$-th and the $\beta$-th molecules are in their electronically excited states, while the other molecules remain in their electronically ground states, i.e., $\ket{\mathrm{E}_{\alpha\beta}} = \hat{\sigma}^{(+)}_\alpha\hat{\sigma}^{(+)}_\beta \ket{\mathrm{G}} $. For the photonic part, the ket state
$\ket{\left\{0\right\}}$ denotes the vacuum state of polaritons (i.e., the zero-polariton state), and the ket state $\ket{ \left\{1_k(\mathbf{r}, \omega)\right\} }$ represents the single-polariton state, i.e., $ \ket{ \left\{1_k(\mathbf{r}, \omega)\right\} } = \hat{f}_k^\dagger (\mathbf{r},\omega)\ket{\left\{0\right\}} $. The coefficients $C^{\mathrm{G},\{0\}}(t)$, $C^{\mathrm{E}_\alpha,\{0\}}(t)$, $C^{\mathrm{G},\left\{1_k\right\}}(\mathbf{r}, \omega, t)$, and $C^{\mathrm{E}_{\alpha \beta},\left\{1_k\right\}}(\mathbf{r}, \omega, t)$ are the probability amplitudes of $\ket{\Psi(t)}$ for the states $\ket{\mathrm{G}} \ket{\left\{0\right\}}$, $\ket{\mathrm{E}_\alpha} \ket{\left\{0\right\}}$, $\ket{\mathrm{G}}\ket{\left\{1_k(\mathbf{r}, \omega)\right\}}$, and
$\ket{ \mathrm{E}_{\alpha \beta}} \ket{\left\{1_k(\mathbf{r}, \omega)\right\} }$, respectively. The symbols $W^{\mathrm{E}_\alpha,\{0\}}$, $W^{\mathrm{G},\{1\}}(\omega)$, and $W^{\mathrm{E}_{\alpha \beta},\{1\}}(\omega)$ correspond to the total energies of the states $\ket{\mathrm{E}_\alpha} \ket{\left\{0\right\}}$, $\ket{\mathrm{G}}\ket{\left\{1_k(\mathbf{r}, \omega)\right\}}$ and
$\ket{ \mathrm{E}_{\alpha \beta}} \ket{\left\{1_k(\mathbf{r}, \omega)\right\} }$, respectively.

To investigate multichromophoric systems coupled with polaritons from vacuum fluctuations, how to properly describe the intermolecular interactions is a key issue. In most cases, rotating interactions (the interactions between 1-excitations $\ket{\mathrm{E}_\alpha} \ket{\left\{0\right\}} $ and $\ket{\mathrm{G}}\ket{\left\{1_k(\mathbf{r}, \omega)\right\}}$) are supposed to be enough to describe multichromophoric systems coupled with polaritons. However, according to the previous study\cite{chuang_2024}, the missing of counter-rotating interactions (the interactions between 1-excitations and 3-excitations $\ket{ \mathrm{E}_{\alpha \beta}} \ket{\left\{1_k(\mathbf{r}, \omega)\right\} }$) result in the absence of the Casimir-Polder potentials of the ground-state molecules and a huge deviation of the intermolecular dipole-dipole interactions. Even in the weak light-matter coupling regime, the deviation of intermolecular dipole-dipole interaction can reach up to 50 percent at a short intermolecular distance. The 0-excitation $\ket{\mathrm{G}} \ket{\left\{0\right\}}$ is used for modeling the total ground state. Note that excitations do not interact with 2-excitations so we neglect these terms in the wavefunction ansatz. As a result, we include basis states 0-, 1-, and 3-excitations into the wavefunction ansatz in Eq.~(\ref{Eq:State_Vector}) in order to better describe dipole-dipole interactions between two molecules\cite{chuang_2024,svidzinsky_2010}. 

By substituting the state vector $\ket{\Psi(t)}$ into the time-dependent Schr\"{o}dinger equation $i\hbar\partial \ket{\Psi(t)}/\partial t = \hat{H}_\mathrm{T} \ket{\Psi(t)}$, we can derive a set of differential equations for the coefficients $C^{\mathrm{E}_\alpha,\{0\}}(t) $ based on our previous work\cite{chuang_2024}. These coefficients are essential for evaluating the LF in Eq.~(\ref{Eq:Lineshape_Func}), given by:
\begin{align}
    \begin{aligned}
\frac{{d}C^{\mathrm{E}_\alpha,\{0\}}(t) }{{d} t} = & -\int_0^t {~d} t^{\prime}  {K}_{\alpha\alpha} (t-t')  e^{i\omega_\alpha (t-t') } C^{\mathrm{E}_\alpha,\{0\}}\left(t^{\prime}\right) \\
& -\sum_{\beta \neq \alpha} \int_0^t {~d} t^{\prime}  {K}_{\alpha\beta} (t-t')   e^{i\omega_\alpha t} e^{-i\omega_\beta t'} C^{\mathrm{E}_\beta,\{0\}}\left(t^{\prime}\right)  \\
& -\sum_{\beta \neq \alpha} \int_0^t {~d} t^{\prime} {K}_{\beta\beta} (t-t')  e^{-i\omega_\beta(t-t')} C^{\mathrm{E}_\alpha,\{0\}}\left(t^{\prime}\right)  \\
& -\sum_{\beta \neq \alpha} \int_0^t {~d} t^{\prime} {K}_{\beta\alpha} (t-t') e^{-i\omega_\beta t}e^{i\omega_a t'}C^{\mathrm{E}_\beta,\{0\}}\left(t^{\prime}\right), 
\label{Eq:ODE_Equation}
\end{aligned}
\end{align}
with the memory kernel of the polariton ${K}_{ab}(\tau)$:
\begin{align}
      {K}_{ab} (\tau) &=  \int_0^\infty d\omega \, \frac{\omega^2}{\pi\hbar \epsilon_0 c^2} {\pmb{\mu}}_a \cdot  \mathrm{Im}\overline{\overline{\mathbf{G}}}(\mathbf{r}_a,\mathbf{r}_b,\omega) \cdot {\pmb{\mu}}_b e^{-i\omega\tau},
\label{Eq:memory_kernel_Polariton}
\end{align}
where $a,b=\alpha\;\mathrm{or}\;\beta$. Additionally, when $a=b$, ${K}_{ab} (\tau)$ is related to $K_\mathrm{pol}(t,t')$ as defined in our previous work\cite{Wang2019}. The dyadic Green's function in Eq.~(\ref{Eq:memory_kernel_Polariton}) can be decomposed into two parts: 
\begin{align}
\label{Eq:Dyadic_Green_Func_Exact}
\overline{\overline{\mathbf{G}}}(\mathbf{r}_a,\mathbf{r}_b,\omega) = \overline{\overline{\mathbf{G}}}_0(\mathbf{r}_a,\mathbf{r}_b,\omega) + \overline{\overline{\mathbf{G}}}_\mathrm{sc}(\mathbf{r}_a,\mathbf{r}_b,\omega),
\end{align}
where $\overline{\overline{\mathbf{G}}}_0(\mathbf{r}_a,\mathbf{r}_b,\omega)$ is the free-space dyadic Green's function, which describes
the direct propagation of photons from $\mathbf{r}_b$ to $\mathbf{r}_a$, and is given by: 
\begin{align}
    \nonumber \overline{\overline{\mathbf{G}}}_0(\mathbf{r}_a,\mathbf{r}_b,\omega)=&
    \frac{e^{ik_0R}}{4\pi R}
    \left\{\vphantom{\frac{e^R}{R}}
    \left(\overline{\overline{\mathbf{I}}}_3-\hat{\mathbf{e}}_\mathrm{R}\otimes\hat{\mathbf{e}}_\mathrm{R}\right)\right.\\
    &+\left.\left(3\hat{\mathbf{e}}_\mathrm{R}\otimes\hat{\mathbf{e}}_\mathrm{R}-\overline{\overline{\mathbf{I}}}_3\right)\left[\frac{1}{(k_0R)^{2}}-\frac{i}{k_0R}\right]
    \right\},
    \label{Eq:g0}
\end{align}
where $k_0 = \omega/c$ and $\mathbf{r}_a - \mathbf{r}_b = R~\hat{\mathbf{e}}_\mathrm{R}=\mathbf{R}$. The symbol $\overline{\overline{\mathbf{G}}}_\mathrm{sc}(\mathbf{r}_a,\mathbf{r}_b,\omega)$ denotes the scattering dyadic Green's function, which describes the propagation of photons after the interaction with the dielectric environment. Various numerical methods can be employed to calculate this dyadic Green's function, including the finite-difference time-domain\cite{KaneYee1966,Hsu2020,Taflove2005}, finite-element\cite{polycarpou2022,Jin2015}, and boundary-element methods\cite{Brebbia1987,POLJAK2005}, which are suitable for general structures of plasmonic materials. For regular geometric structures, analytical approaches such as the Fresnel method\cite{wu2018characteristic,Wang2020} and Mie theory\cite{Ding2018,Lee2020} can also be applied.

The full dynamical equation Eq.~(\ref{Eq:ODE_Equation}) can be simplified under weak light-matter interactions regime with the Markov approximation, as shown in our previous work\cite{chuang_2024}, resulting in the following expression: 
\begin{align}
\nonumber
\frac{{d C^{\mathrm{E}_\alpha,\{0\}}(t)}}{{d} t} =&  -\frac{i}{\hbar}\left\{\left[\Delta_{\mathrm{e}_\alpha}+\sum_{\beta \neq \alpha} \Delta_{\mathrm{g}_\beta}\right]-i \hbar \frac{\Gamma_\alpha}{2}\right\} C^{\mathrm{E}_\alpha,\{0\}}(t) \\
& -\frac{i}{\hbar} \sum_{\beta \neq \alpha} \mathrm{V}_{\mathrm{DDI}, \alpha \beta} e^{-i\left(\omega_\beta-\omega_\alpha\right) t} C^{\mathrm{E}_\beta,\{0\}}(t),
\label{Eq:Analytical_Solution_Dynamics}
\end{align}
where $\Delta_{\mathrm{e(g)}_{\alpha(\beta)}}$ represents the energy shift of the excited (ground) state of the $\alpha (\beta)$-th molecule, which includes contributions from the free-space Lamb shifts and the Casimir-Polder potentials\cite{buhmann2013dispersion_1,buhmann2013dispersion_2}. In the system that we will discuss later, these energy shifts are negligible. The decay rate $\Gamma_\alpha$ is given by:
\begin{align}
    \Gamma_\alpha = \frac{2\omega_\alpha^2}{\hbar \epsilon_0 c^2} \pmb{\mu}_\alpha \cdot \mathrm{Im} \, \overline{\overline{\mathbf{G}}} (\mathbf{r}_\alpha, \mathbf{r}_\alpha, \omega_\alpha) \cdot \pmb{\mu}_\alpha,
\label{Eq:Decay_Rate_Gamma}
\end{align}
which aligns with the spontaneous emission rate of a molecule in a medium, derived from Fermi’s golden rule\cite{novotny2012principles}. The dipole-dipole interaction $V_{\mathrm{DDI},\alpha\beta}$ can be divided into two components $V_{\mathrm{DDI},\alpha\beta} = V_{\mathrm{RDDI},\alpha\beta} + V_{\mathrm{ORC},\alpha\beta}$, where $V_{\mathrm{RDDI},\alpha\beta}$ represents the resonant dipole-dipole interaction between a pair of molecules\cite{Dung2002_Energy}:
\begin{align}
&V_{\mathrm{RDDI},\alpha\beta} = \frac{-\omega_\beta^2}{\epsilon_0 c^2} \pmb{\mu}_\alpha \cdot \overline{\overline{\mathbf{G}}}(\mathbf{r}_\alpha, \mathbf{r}_\beta, \omega_\beta) \cdot \pmb{\mu}_\beta.\label{Eq:RDDI} 
\end{align}

In the system we study later, where all molecules are identical ($\omega_\alpha = \omega_\beta$ and $\pmb{\mu}_\alpha=\pmb{\mu}_\beta$), the off-resonance dipole-dipole interaction $V_\mathrm{ORC,\alpha\beta} = 0$ and Eq.~(\ref{Eq:Analytical_Solution_Dynamics}) can be expressed in a matrix form with a non-Hermitian Hamiltonian matrix ${\tilde{\mathbf{H}}}^{(N_\mathrm{M})}$:
\begin{align}
    i\hbar\frac{d \mathbf{C}_{N_\mathrm{M}}(t) }{ d t} = {\tilde{\mathbf{H}}}^{(N_\mathrm{M})} \mathbf{C}_{N_\mathrm{M}}(t)
\label{Eq:non_Hermitian_Schr_Eq}
\end{align} 
where $\mathbf{C}_{N_\mathrm{M}}(t) = \left(C^{\mathrm{E}_1,\{0\}}(t), C^{\mathrm{E}_2,\{0\}}(t),\cdots, C^{\mathrm{E}_{N_\mathrm{M}},\{0\}}(t)\right)^\mathrm{T} $ is a column vector containing the coefficients $C^{\mathrm{E}_\alpha,\{0\}}(t)$. The matrix element of the $N_\mathrm{M} \times N_\mathrm{M}$ non-Hermitian Hamiltonian matrix is $  {\tilde{{H}}}^{(N_\mathrm{M})}_{ij} = - i\frac{\hbar\Gamma_i}{2} \delta_{ij} + V_{\mathrm{RDDI},ij}(1-\delta_{ij})$. For example, the non-Hermitian Hamiltonian matrix for the dimer and trimer systems can be expressed as: 
\begin{align}
\label{Eq_H_Eff2}
 &\tilde{\mathbf{H}}^{(2)} = \begin{bmatrix}
         - i{\hbar\Gamma_1}/{2} & V_{\mathrm{RDDI},12} \\
          V_{\mathrm{RDDI},21} & - i{\hbar\Gamma_2}/{2} \\
    \end{bmatrix}, \\
 &\tilde{\mathbf{H}}^{(3)} = \begin{bmatrix}
           - i{\hbar\Gamma_1}/{2}  & V_{\mathrm{RDDI},12} & V_{\mathrm{RDDI},13} \\
        V_{\mathrm{RDDI},21} &    - i{\hbar\Gamma_2}/{2}  & V_{\mathrm{RDDI},23}\\
        V_{\mathrm{RDDI},31} & V_{\mathrm{RDDI},32} &    - i{\hbar\Gamma_3}/{2}  \\
    \end{bmatrix},
\label{Eq_H_Eff3}
\end{align}
where the resonant dipole-dipole interaction $V_{\mathrm{RDDI},\alpha\beta}= \Re{V_{\mathrm{RDDI},\alpha\beta}}+i\Im{V_{\mathrm{RDDI},\alpha\beta}}$ is a complex number. Additionally, $V_{\mathrm{RDDI},\alpha\beta}=V_{\mathrm{RDDI},\beta\alpha} \neq V^*_{\mathrm{RDDI},\beta\alpha} $ due to the property\cite{Dung1998} of the dyadic Green's function $\overline{\overline{\mathbf{G}}}(\mathbf{r}_\alpha, \mathbf{r}_\beta, \omega) = \left( \overline{\overline{\mathbf{G}}}(\mathbf{r}_\beta, \mathbf{r}_\alpha, \omega) \right)^\mathrm{T} \neq \left( \overline{\overline{\mathbf{G}}}(\mathbf{r}_\beta, \mathbf{r}_\alpha, \omega) \right)^\dagger $. Notably, in a previous work based on QED\cite{spano_1989}, Spano and Mukamel also employed a non-Hermitian Hamiltonian with complex dipole-dipole interactions to investigate the superradiance of molecules in a vacuum. However, to the best of our knowledge, no prior work has combined a non-Hermitian Hamiltonian with emission power spectra to study the spontaneous emission of molecules in inhomogeneous environments, as we do in this work.

In summary, Eq.~(\ref{Eq:ODE_Equation}) describes the quantum dynamics of a multichromophoric system in a complex dielectric environment, applicable in both strong and weak light-matter coupling regimes. Eqs.~(\ref{Eq:Analytical_Solution_Dynamics}) and (\ref{Eq:non_Hermitian_Schr_Eq}) derived from Eq.~(\ref{Eq:ODE_Equation}) are specific in the weak light-matter coupling regime under the Markov approximation. Notably, all these quantum dynamical equations incorporate the effects of counter-rotating interactions.

\subsection{Theory of Emission Power Spectra}

\subsubsection{Analytical Expression of Emission Power Spectra}
\label{Sec_Formalism_EPS}

The emission power spectrum $ \mathcal{S}^{(N_\mathrm{M})}(\mathbf{R}_\mathrm{D},\omega_\mathrm{D})$ of a collection of $N_\mathrm{M}$ molecules coupled to polaritons in the arbitrary dielectric medium can be defined in terms of the two-time correlation function of the field operator as follows:\cite{Eberly1977,scully_1997,holdaway2018perturbation,Dung2000} 
\begin{align}
\label{Eq:Definiation_Power_Spectrum_Formal} 
    \mathcal{S}^{(N_\mathrm{M})}(\mathbf{R}_\mathrm{D},\omega_\mathrm{D})&\equiv \int_0^\infty dt_2 \int_0^\infty dt_1 e^{-i\omega_\mathrm{D}(t_2-t_1)} {C}(\mathbf{R}_\mathrm{D},t_1,t_2).
\end{align}
This formula is exact and not constrained by the truncated state vector, where $\omega_\mathrm{D}$ is the frequency of the detected photon, and ${C}(\mathbf{R}_\mathrm{D},t_1,t_2)$ is the two-time correlation function of the field operator at the detector position $\mathbf{R}_\mathrm{D}$, defined in the Heisenberg picture as: 
\begin{align}
\label{Eq:Time-Correlation_Def}
    {C}(\mathbf{R}_\mathrm{D},t_1,t_2) &= \left<\hat{\mathbf{E}}^{(-)}(\mathbf{R}_\mathrm{D},t_2)\cdot \hat{\mathbf{E}}^{(+)}(\mathbf{R}_\mathrm{D},t_1)\right>,
\end{align}
where $\hat{\mathbf{E}}^{(+)}(\mathbf{R}_\mathrm{D},t_1)=e^{i\hat{H}_\mathrm{T}t_1/\hbar}\hat{\mathbf{E}}^{(+)}(\mathbf{R}_\mathrm{D})e^{-i\hat{H}_\mathrm{T}t_1/\hbar}$ and $\hat{\mathbf{E}}^{(-)}(\mathbf{R}_\mathrm{D},t_2) = e^{i\hat{H}_\mathrm{T}t_2/\hbar}\hat{\mathbf{E}}^{(-)}(\mathbf{R}_\mathrm{D})e^{-i\hat{H}_\mathrm{T}t_2/\hbar}$ are the Heisenberg picture representations of $\hat{\mathbf{E}}^{(+)}(\mathbf{R}_\mathrm{D})$ and $\hat{\mathbf{E}}^{(-)}(\mathbf{R}_\mathrm{D})$, respectively. Inspired by previous works\cite{medina_2021,chuang_2024_2}, the equation of motion for $\hat{\mathbf{E}}^{(+)}(\mathbf{R}_\mathrm{D},t_1)$ can be derived (see Appendix \ref{Appendix_Eq:E(+)_Integral_Form}
for the details):
\begin{align}
\nonumber
    &\hat{\mathbf{E}}^{(+)}(\mathbf{R}_\mathrm{D},t_1)\\&= \hat{\mathbf{E}}^{(+)}_\mathrm{free}(\mathbf{R}_\mathrm{D},t_1)  + i \sum_{\alpha=1}^{N_\mathrm{M}} \int_0^{t_1} dt' \;  \mathbf{K}_\alpha (\mathbf{R}_\mathrm{D},t_1-t') 
   \hat{\bar{\mu}}_\alpha(t'),
\label{Eq:E(+)_Integral_Form}
\end{align}
where $\hat{\mathbf{E}}^{(+)}_\mathrm{free}(\mathbf{R}_\mathrm{D},t_1)=\int \mathrm{d}\mathbf{r}   \int_0^\infty d\omega \,  e^{-i\omega t_1}\, \overline{\overline{\mathbf{g}}}(\mathbf{R}_\mathrm{D},\mathbf{r},\omega) \cdot\hat{\mathbf{f}}(\mathbf{r},\omega,0)$ describes the free evolution of $\hat{\mathbf{E}}^{(+)}(\mathbf{R}_\mathrm{D},t_1)$. The memory kernel $\mathbf{K}_\alpha (\mathbf{R}_\mathrm{D},\tau)$ in Eq.~(\ref{Eq:E(+)_Integral_Form}) is related to the propagation of emitted photons from the $\alpha$-th molecule to the detector:
\begin{align}
\label{Eq:Memory_Kernel_Propergation}
      \mathbf{K}_\alpha (\mathbf{R}_\mathrm{D},\tau) &=  \int_0^\infty d\omega \, \frac{\omega^2}{\pi\epsilon_0 c^2}  \mathrm{Im}\overline{\overline{\mathbf{G}}}(\mathbf{R}_\mathrm{D},\mathbf{r}_\alpha,\omega) \cdot {\pmb{\mu}}_\alpha e^{-i\omega\tau}. 
\end{align}
Note that $\mathbf{K}_\alpha (\mathbf{R}_\mathrm{D},\tau)$ in Eq.~(\ref{Eq:Memory_Kernel_Propergation}) is a $3\times 1$ vector function related to the EEF, which is fundamentally different from the scalar function ${K}_{ab}(\tau)$ in Eq.~(\ref{Eq:memory_kernel_Polariton}) associated with the LF. Since $\hat{\mathbf{E}}^{(-)}(\mathbf{R}_\mathrm{D},t_2)$ is the Hermitian conjugate of $\hat{\mathbf{E}}^{(+)}(\mathbf{R}_\mathrm{D},t_2)$ as shown in Eq.~(\ref{Eq:E(-)}), and we are studying the spontaneous emission process (i.e., the polariton field is initially in the vacuum state), we can simplify Eq.~(\ref{Eq:Time-Correlation_Def}) (see Appendix \ref{Appendix_Eq:Correlation_Express_2}
for the details) as:
\begin{align}
\nonumber
    {C}(\mathbf{R}_\mathrm{D},t_1,t_2) =& \sum_{\alpha,\beta=1}^{N_\mathrm{M}} \int_0^{t_1} dt' \int_0^{t_2} dt'' \; \left<   \hat{\bar{\mu}}_\beta(t'') 
   \hat{\bar{\mu}}_\alpha(t') \right>\\
    &\times \mathbf{K}^*_\beta (\mathbf{R}_\mathrm{D},t_2-t'') \cdot    \mathbf{K}_\alpha (\mathbf{R}_\mathrm{D},t_1-t') .
\label{Eq:Correlation_Express_2}
\end{align} 

Note that Eq.~(\ref{Eq:Correlation_Express_2}) can be reduced to the Eq.~(10) in J. Feist \textit{et al.}'s work\cite{medina_2021} under the conditions $\alpha = \beta = N_\mathrm{M} = 1$ (representing one molecule) with some differences in forms. Eq.~(\ref{Eq:Correlation_Express_2}) represents the two-time correlation function of the field operator, whereas the Eq.~(10) in J. Feist \textit{et al.}'s work\cite{medina_2021} gives the electric field intensity at a specific time $t$. These differences arise from the two Laplace transforms in Eq.~(\ref{Eq:Definiation_Power_Spectrum_Formal}). Substituting Eq.~(\ref{Eq:Correlation_Express_2}) into Eq.~(\ref{Eq:Definiation_Power_Spectrum_Formal}), and after some algebra (the details provided in Appendix \ref{Appendix_EPS_EXact}), we find that the emission power spectrum $\mathcal{S}^{(N_\mathrm{M})}(\mathbf{R}_\mathrm{D},\omega_\mathrm{D}) $ can be written as a sum of products
of the electromagnetic
environment factors (EEF) $\mathcal{F}^{(N_\mathrm{M})}_{\alpha,\beta}(\mathbf{R}_\mathrm{D},\omega_\mathrm{D}) $ and the lineshape functions (LF) $\mathcal{L}^{(N_\mathrm{M})}_{\alpha,\beta}(\omega_\mathrm{D})$ as follows:
\begin{align}
\label{Eq:EPS_Exact}
     \mathcal{S}^{(N_\mathrm{M})}(\mathbf{R}_\mathrm{D},\omega_\mathrm{D})   =&\sum_{\alpha,\beta=1}^{N_\mathrm{M}} \mathcal{L}^{(N_\mathrm{M})}_{\alpha,\beta}(\omega_\mathrm{D}) \mathcal{F}^{(N_\mathrm{M})}_{\alpha,\beta}(\mathbf{R}_\mathrm{D},\omega_\mathrm{D})      , 
\end{align}
with
\begin{align}
&\mathcal{L}^{(N_\mathrm{M})}_{\alpha,\beta}(\omega_\mathrm{D}) =  \int_0^\infty dt'' \int_0^{\infty} dt' \; e^{-i\omega_\mathrm{D}(t''-t')}  
 \left< \hat{\bar{\mu}}_\beta(t'') 
   \hat{\bar{\mu}}_\alpha(t') \right> ,
\label{Eq:Lineshape_Func} \\
\nonumber
    &\mathcal{F}^{(N_\mathrm{M})}_{\alpha,\beta}(\mathbf{R}_\mathrm{D},\omega_\mathrm{D})\\
    &=\int_{0}^\infty d\tau_1 \int_{0}^\infty d\tau_2 \; e^{-i\omega_\mathrm{D}(\tau_2-\tau_1)}  \mathbf{K}^*_\beta (\mathbf{R}_\mathrm{D},\tau_2) \cdot    \mathbf{K}_\alpha (\mathbf{R}_\mathrm{D},\tau_1).
\label{Eq:EEF_Exact}
\end{align}
\normalsize 

Note that Eq.~(\ref{Eq:EPS_Exact}) provides the analytical formula for the emission power spectra of a collection of molecules, which is NOT constrained by our truncated state vector in Eq.~(\ref{Eq:State_Vector}). The analytical formalism of the LF in Eq.~(\ref{Eq:Lineshape_Func}) is expressed in terms of the two-time correlation function of $\hat{\bar{\mu}}_{\alpha(\beta)}$, while the analytical formalism of the EEF in Eq.~(\ref{Eq:EEF_Exact}) is given in terms of the memory kernel $\mathbf{K}_{\alpha(\beta)} (\mathbf{R}_\mathrm{D},\tau)$ (recall Eq.~(\ref{Eq:Memory_Kernel_Propergation})).

\subsubsection{Approximate Expression of Emission Power Spectra based on Truncated State Vector}
\label{Sec_Approx_EPS}

The two-time correlation function $\left< \hat{\bar{\mu}}_\beta(t'') 
   \hat{\bar{\mu}}_\alpha(t') \right> $ in the definition of the LF (see Eq.~(\ref{Eq:Lineshape_Func})) can be evaluated based on the truncated state vector in Eq.~(\ref{Eq:State_Vector}) as follows:
\begin{align}
\nonumber
    \left< \hat{\bar{\mu}}_\beta(t'') 
   \hat{\bar{\mu}}_\alpha(t') \right> &=  \bra{\Psi(0)}  \hat{\bar{\mu}}_\beta(t'') 
   \hat{\bar{\mu}}_\alpha(t') \ket{\Psi(0)} \\
   &=\bra{\Psi(t'')} \hat{\bar{\mu}}_\beta
    e^{\frac{i\hat{H}_\mathrm{T}(t'-t'')}{\hbar}} \hat{\bar{\mu}}_\alpha \ket{\Psi(t')}.
\label{Eq:Two_time_Schrodinger}
\end{align}

Substituting Eq.~(\ref{Eq:Two_time_Schrodinger}) into Eq.~(\ref{Eq:Lineshape_Func}) and after some algebra (see Appendix \ref{Appendix_Time_Correlation_Functino_Dipole}
for details), we obtain the expression for the LF evaluated based on the truncated state vector as follows:  
\begin{align}
\nonumber \mathcal{L}^{(N_\mathrm{M})}_{\alpha,\beta}(\omega_\mathrm{D})  
   \approx & \left[ \mathfrak{L}_{t''} \left\{C^{\mathrm{E}_{\beta},\{0\}}(t'') \right\}(i(W^{\mathrm{E}_{\beta},\{0\}}-\omega_\mathrm{D}) ) \right]^* \\
& \times \mathfrak{L}_{t'} \left\{C^{\mathrm{E}_{\alpha},\{0\}}(t') \right\}(i(W^{\mathrm{E}_{\alpha},\{0\}}-\omega_\mathrm{D}) ) ,
\label{Eq:Two-Time_Correlation_App}
\end{align}
where $\mathfrak{L}_\tau\left\{f(\tau)\right\}(s)=\int_0^\infty d\tau e^{-s\tau} f(\tau)$ denotes the Laplace transform of a function $f(\tau)$ at a variable $s$. The coefficients $C^{\mathrm{E}_{\alpha},\{0\}}(t')$ and $C^{\mathrm{E}_{\beta},\{0\}}(t')$ are obtained by simulating the full quantum dynamical equations in Eq.~(\ref{Eq:ODE_Equation}). Equation~(\ref{Eq:Two-Time_Correlation_App}) provides an approximate expression for the LF within the truncated state space, while remaining applicable to multichromophoric systems in arbitrary inhomogeneous, dispersive, and absorbing media.

By adopting the notation for the Laplace transform, $\mathfrak{L}_\tau\left\{f(\tau)\right\}(s)$, we can rewrite the analytical expression for the EEF in Eq.~(\ref{Eq:EEF_Exact}) as follows:
\begin{align} 
\nonumber
    &\mathcal{F}^{(N_\mathrm{M})}_{\alpha,\beta}(\mathbf{R}_\mathrm{D},\omega_\mathrm{D})\\
    &=\mathfrak{L}_{\tau_2}\left\{ \mathbf{K}^*_\beta (\mathbf{R}_\mathrm{D},\tau_2) \right\}(i\omega_\mathrm{D}) \cdot \mathfrak{L}_{\tau_1}\left\{ \mathbf{K}_\alpha (\mathbf{R}_\mathrm{D},\tau_1) \right\}(-i\omega_\mathrm{D}).
\label{Eq:GEEF_Alternative_Form}
\end{align}
Equation~(\ref{Eq:GEEF_Alternative_Form}) can be reduced to the following form based on Kramers-Kronig relations\cite{lee_2021} (see Appendix \ref{Appendix_Eq:GEEF_Reduced} for the details): 
\begin{align}
\nonumber
    &\mathcal{F}^{(N_\mathrm{M})}_{\alpha,\beta}(\mathbf{R}_\mathrm{D},\omega_\mathrm{D})
    \\ &\approx \left( \frac{\omega_\mathrm{D}^2}{\epsilon_0 c^2} \right)^2 {\boldsymbol{\mu}}_\alpha \cdot \overline{\overline{\mathbf{G}}}^\dagger(\mathbf{R}_\mathrm{D},\mathbf{r}_\alpha,\omega_\mathrm{D})  \cdot   \overline{\overline{\mathbf{G}}}(\mathbf{R}_\mathrm{D},\mathbf{r}_\beta,\omega_\mathrm{D}) \cdot {\boldsymbol{\mu}}_\beta .
\label{Eq:GEEF_Reduced}
\end{align}

In conclusion, Eqs.~(\ref{Eq:EPS_Exact}), (\ref{Eq:Two-Time_Correlation_App}), and (\ref{Eq:GEEF_Reduced}) are the main formulas for calculating
the emission power spectrum of a collection of excited molecules in arbitrary inhomogeneous, dispersive, and absorbing media, assuming that the truncated state vector provides a valid representation of the light-matter hybrid system.

\subsection{Specific Expression of Emission Power Spectra for Planar Systems}
\label{Sec:EPS_Planar}

\begin{figure}[htbp] \includegraphics[width=0.48\textwidth]{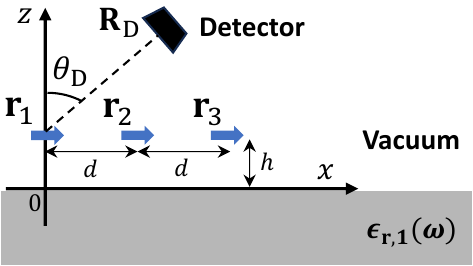}
\caption{Schematic illustration of the setup for measuring the emission power spectra of a collection of molecules above a silver surface. Three identical molecules (indicated by blue arrows) are positioned at $\mathbf{r}_1$, $\mathbf{r}_2$, and $\mathbf{r}_3$. The molecular transition dipole moments are aligned along the $x$-axis. The parameters $d$ and $h$ represent the intermolecular distance and the height above the silver surface, respectively. $\mathbf{R}_\mathrm{D}$ denotes the position of the detector, and $\theta_\mathrm{D}$ is the
angle between the $z$-axis and the dashed line connecting the first molecule to the detector.}
\label{Fig_2}
\end{figure}

In this section, we focus on the specific expression of the EPS for a collection of molecules placed above a planar surface. The setup, illustrated in Fig.~\ref{Fig_2}, involves up to three identical molecular emitters arranged in a linear chain with an intermolecular distance $d$ and at a height of $h$ above the planar surface. Specifically, the three molecules are positioned at $\mathbf{r}_1=(0,0,h)$, $\mathbf{r}_2=(d,0,h)$, and $\mathbf{r}_3=(2d,0,h)$, with the detector located at $\mathbf{R}_\mathrm{D}=(x,y,z)$. The planar surface can be modeled via the following dielectric functions:
\begin{align}
    \epsilon_\mathrm{r}(\mathbf{r},\omega) = 
    \begin{cases}
        \epsilon_{\mathrm{r},0}(\omega)=1,     & \text{ if } z>0, \\
         \epsilon_{\mathrm{r},1}(\omega),    & \text{ if } z<0,
    \end{cases} 
    \label{dielectrics}
\end{align}
where $\epsilon_{\mathrm{r},0}(\omega)=1$ is the dielectric constant of vacuum, and $\epsilon_{\mathrm{r},1}(\omega)$ represents the the dielectric function of a plasmonic material.  

In this work, we focus on the EEF in the far-field region, where the distance between the first molecule and the detector is much larger than the wavelength of the detected photon, i.e., $\abs{\mathbf{R}_\mathrm{D}-\mathbf{r}_{1}}\gg \lambda_\mathrm{D} = 2\pi c/\omega_\mathrm{D}$. According to our previous study\cite{lee_2021}, the dyadic Green's function $\overline{\overline{\mathbf{G}}}(\mathbf{R}_\mathrm{D},\mathbf{r}_1,\omega_\mathrm{D})$ in Eq.~(\ref{Eq:GEEF_Reduced}), which describes the propagation of photons from the first molecule to the detector, can be approximated in the far-field region as $\overline{\overline{\mathbf{G}}}_\mathrm{far}(\mathbf{R}_\mathrm{D},\mathbf{r}_1,\omega_\mathrm{D})$:
\begin{align}
\nonumber      \overline{\overline{\mathbf{G}}}_\mathrm{far}(\mathbf{R}_\mathrm{D},\mathbf{r}_1,\omega_\mathrm{D}) = \, &\overline{\overline{\mathbf{G}}}_{0,\mathrm{far}}(\mathbf{R}_\mathrm{D},\mathbf{r}_1,\omega_\mathrm{D}) +   \overline{\overline{\mathbf{G}}}_{\mathrm{s,far}}(\mathbf{R}_\mathrm{D},\mathbf{r}_1,\omega_\mathrm{D}) \\
    &+\overline{\overline{\mathbf{G}}}_{\mathrm{p,far}}(\mathbf{R}_\mathrm{D},\mathbf{r}_1,\omega_\mathrm{D}) , 
\label{Eq:Total_Greens_Func}
\end{align}
where $\overline{\overline{\mathbf{G}}}_{0,\mathrm{far}}(\mathbf{R}_\mathrm{D},\mathbf{r}_1,\omega_\mathrm{D})$ and $\overline{\overline{\mathbf{G}}}_{\mathrm{s(p),far}}(\mathbf{R}_\mathrm{D},\mathbf{r}_1,\omega_\mathrm{D})$ are the free-space and the s-polarized (p-polarized) dyadic Green's function in the far-field region, respectively. The free-space dyadic Green's function in the far-field region is given by:
\begin{align}
\label{Eq:Green_Func_Vac}
    \tensorg_{0,\mathrm{far}}(\mathbf{R}_\mathrm{D},\mathbf{r}_1,\omega_\mathrm{D})=&
    \frac{e^{ik_0R}}{4\pi R}
    \left(\overline{\overline{\mathbf{I}}}_3-\hat{\mathbf{e}}_{R}\otimes\hat{\mathbf{e}}_{R}\right) ,
\end{align}
where $R = \abs{\mathbf{R}_\mathrm{D} - \mathbf{r}_1 } = \sqrt{ x^2+y^2+(z-h)^2 }$ is the distance between the first molecule and the detector, $\hat{\mathbf{e}}_{R} = \left( \mathbf{R}_\mathrm{D} - \mathbf{r}_1 \right) /R = (x,y,z-h)/\sqrt{ x^2+y^2+(z-h)^2 } $ is the corresponding unit vector, and $k_0=\omega_\mathrm{D}/c$ is the magnitude of wave vector in vacuum. The s-polarized (or p-polarized) dyadic Green's function in the far-field region can be expressed as:
\begin{align}
    \tensorg_{\sigma,\mathrm{far}}(\mathbf{R}_\mathrm{D},\mathbf{r}_1,\omega_\mathrm{D}) &=
    \frac{e^{ik_0\bar{R}}}{4\pi\bar{R}}r_{\sigma,01}(\bar{q},\omega_\mathrm{D})
    \overline{\overline{\mathbf{M}}}_{\sigma\mathrm{,far}}(\bar\theta,\bar\phi),
    \label{Eq:Gsigmafar}
\end{align}
where $\sigma = \mathrm{s\;or\;p}$, $\bar{R} = \sqrt{x^2+y^2+(z+h)^2}$ is the distance between the detector and the mirror dipole positioned at $(0,0,-h)$. The term $r_{\sigma,01}(\bar{q},\omega_\mathrm{D})$ denotes the $\sigma$-polarized reflection coefficient between vacuum and silver, and $\overline{\overline{\mathbf{M}}}_{\sigma,\mathrm{far}}(\bar\theta,\bar\phi)$ denotes the $\sigma$-polarized auxiliary tensor in the far-field region. The detailed expression of $r_{\sigma,01}(\bar{q},\omega_\mathrm{D})$ and $\overline{\overline{\mathbf{M}}}_{\sigma,\mathrm{far}}(\bar\theta,\bar\phi)$ can be found in Eqs.~(\ref{App_Eq_Reflection_Coeff})-(\ref{Eq:M_p_far}) in Appendix~\ref{App_Reflection_FarField}.

The dyadic Green's function $\overline{\overline{\mathbf{G}}}(\mathbf{R}_\mathrm{D},\mathbf{r}_{2(3)},\omega_\mathrm{D})$ in Eq.~(\ref{Eq:GEEF_Reduced}), which describes the propagation of a photon from the second (or third) molecule to the detector, can be approximated by $\overline{\overline{\mathbf{G}}}(\mathbf{R}_\mathrm{D},\mathbf{r}_1,\omega_\mathrm{D}) $ when $\abs{\mathbf{r}_3-\mathbf{r}_1}=2d \ll \rho $ and $ 2d \ll \lambda_\mathrm{D}  $, where $\rho= \sqrt{x^2+y^2}$ is the radial distance of the detector. The validity of this approximation is proven in Appendix \ref{App_EEF_r2}. Consequently, the EEFs given in Eq.~(\ref{Eq:GEEF_Reduced}) for closely aligned molecular aggregates can be simplified to:
\begin{align}
\mathcal{F}^{(N_\mathrm{M})}_{\alpha,\beta}(\mathbf{R}_\mathrm{D},\omega_\mathrm{D}) \approx \mathcal{F}^{(N_\mathrm{M})}_{1,1}(\mathbf{R}_\mathrm{D},\omega_\mathrm{D}), 
\label{Eq:EEFs_for_Closely_Aligned}
\end{align}
for any $\alpha,\beta=1,2,3$.  Substituting Eq.~(\ref{Eq:EEFs_for_Closely_Aligned}) into the general theory for the EPS in Eq.~(\ref{Eq:EPS_Exact}), we obtain the specific EPS for the planar system as follows:
\begin{align}
     \mathcal{S}^{(N_\mathrm{M})}(\mathbf{R}_\mathrm{D},\omega_\mathrm{D})   \approx & \mathcal{L}^{(N_\mathrm{M})}(\omega_\mathrm{D}) \mathcal{F}^{(N_\mathrm{M})}_{1,1}(\mathbf{R}_\mathrm{D},\omega_\mathrm{D}),
     \label{Eq:Plane_EPS}
\end{align}
with
\begin{align}
\label{Eq:LF_11}
   &\mathcal{L}^{(N_\mathrm{M})}(\omega_\mathrm{D}) = \sum_{\alpha,\beta=1}^{N_\mathrm{M}} \mathcal{L}^{(N_\mathrm{M})}_{\alpha,\beta}(\omega_\mathrm{D}),\\
\label{Eq:EEF_11}
    &\mathcal{F}^{(N_\mathrm{M})}_{1,1}(\mathbf{R}_\mathrm{D},\omega_\mathrm{D}) \simeq  \abs{ \frac{\omega_\mathrm{D}^2}{\epsilon_0 c^2}       \overline{\overline{\mathbf{G}}}_\mathrm{far}(\mathbf{R}_\mathrm{D},\mathbf{r}_1,\omega_\mathrm{D}) \cdot {\pmb{\mu}}_1 }^2, 
\end{align}
where the far-field EEF $\mathcal{F}^{(N_\mathrm{M})}_{1,1}(\mathbf{R}_\mathrm{D},\omega_\mathrm{D})$ in Eq.~(\ref{Eq:EEF_11}) utilizes the (far-field) dyadic Green's functions in Eqs.~(\ref{Eq:Total_Greens_Func})-(\ref{Eq:Gsigmafar}) for the planar system, describing photon propagation from the molecule to the detector located at $\mathbf{R}_\mathrm{D}$. In contrast, the LF based on wavefunction ansatz $\mathcal{L}^{(N_\mathrm{M})}(\omega_\mathrm{D})$ in Eq.~(\ref{Eq:LF_11}) is the sum of $\mathcal{L}^{(N_\mathrm{M})}_{\alpha,\beta}(\omega_\mathrm{D})$ from Eq.~(\ref{Eq:Two-Time_Correlation_App}), where $C^{\mathrm{E}_{\alpha},\{0\}}(t')$ is calculated using the quantum dynamical equations in Eq.~(\ref{Eq:ODE_Equation}) and the exact dyadic Green's function for a pair of molecules in Eq.~(\ref{Eq:Dyadic_Green_Func_Exact}). Note that the EPS in Eq.~(\ref{Eq:Plane_EPS}) is written as a single product of the far-field EEF and the LF based on wavefunction ansatz, which is valid for closely aligned molecular aggregates ($2d\ll \lambda_\mathrm{D}$), located far from the detector ($2d \ll \rho$), and in the far-field region ($\left|\mathbf{R}_\mathrm{D}-\mathbf{r}_1  \right| \gg \lambda_\mathrm{D}$).

\begin{widetext}
    \begin{figure*}[htbp] \includegraphics[width=0.96\textwidth]{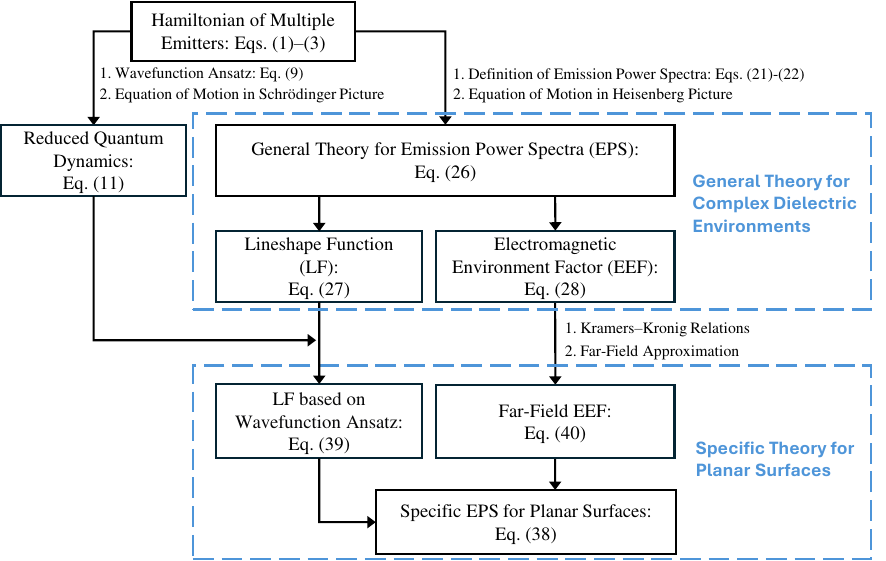}
\caption{Theoretical structure of molecular emission power spectra for multichromophoric systems: the relationship between the general theory and the specific theory for planar surfaces.}
\label{flowchart}
\end{figure*}
\end{widetext}

To conclude, Eqs.~(\ref{Eq:Plane_EPS})-(\ref{Eq:EEF_11}) are the main formulas for calculating
the EPS of a closely aligned molecular aggregate above a planar system. We also provide a summary of the relationship between the general theory of EPS for arbitrary dielectric environments and the specific theory of EPS for planar surfaces, as outlined in Fig.~\ref{flowchart}. 

\section{Numerical Demonstration and Discussion}
\label{Sec:Discussion}

\subsection{LFs of Molecules above a Silver Surface}
\label{SubSec:LF}

\begin{figure}[htbp] \includegraphics[width=0.5\textwidth]{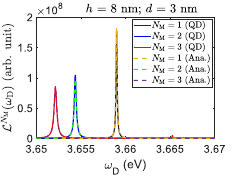}
\caption{Frequency dependence of the LFs of molecular emitters oriented along the $x$-axis with the intermolecular distance of $d=3$ nm and at the height of $h=8$ nm above the silver surface. The solid lines represent the LFs obtained from numerical quantum dynamics simulations (QD), while the dashed lines represent the analytical (Ana.) solutions. Different values of $N_\mathrm{M}$ indicate the number of molecules, as shown in the legend.}
\label{Fig_4}
\end{figure}

In this section, we numerically demonstrate the LFs for molecular aggregates consisting of one to three molecules placed in the environment depicted in Fig~\ref{Fig_2}. Specifically, we selected an intermolecular distance of $d=3.0$ nm to represent weak dipole-dipole interactions and positioned the molecules $h=8.0$ nm above the silver surface to ensure weak light-matter coupling. The dielectric functions of silver, $\epsilon_\mathrm{r,1}(\omega)=\epsilon_\mathrm{Ag}(\omega)$, is obtained by fitting experimental data\cite{Johnson1972,Li1976} with a sophisticated model\cite{Melikyan2014} that ensures Kramers-Kronig relations\cite{scheel_1998}. In our simulation, we focus on the spontaneous emission of a group of molecules in their delocalized exciton states, so the initial condition is set as $C^{\mathrm{E}_\alpha,\{0\}}(t=0) = 1/\sqrt{N_\mathrm{M}}$, with the transition frequency $\hbar\omega_\alpha = 3.659$ eV and the transition dipole moment magnitude $\abs{\pmb{\mu}_\alpha} =10$ Debye for $\alpha = 1,\cdots,N_\mathrm{M}$. For instance, in the dimer system ($N_\mathrm{M}=2$), the initial condition becomes $C^{\mathrm{E}_1,\{0\}}(t=0) =C^{\mathrm{E}_2,\{0\}}(t=0) = 1/\sqrt{2}$, with both transition frequencies set to $\hbar\omega_1=\hbar\omega_2 = 3.659$ eV and transition dipole moment magnitudes $\abs{\pmb{\mu}_1} = \abs{\pmb{\mu}_2} =10$ Debye. The quantum dynamics of coefficients $C^{\mathrm{E}_\alpha,\{0\}}(t)$ in Eq.~(\ref{Eq:ODE_Equation}) are calculated to evaluate the LF in Eq.~(\ref{Eq:LF_11}), and are plotted in Fig.~\ref{Fig_4} referred to as QD.

Figure~\ref{Fig_4} illustrates the LFs in the regime of weak light-matter interactions, highlighting three key phenomena: (i) the peak positions redshift as the number of molecules increases, with $N_\mathrm{M} = 3$ (red) exhibiting the largest shift compared to $N_\mathrm{M} = 2$ (blue) and $N_\mathrm{M} = 1$ (black); (ii) the peak width broadens with more molecules, as indicated by the increasing full width at half-maximum (FWHM) from $N_\mathrm{M} = 1$ to $N_\mathrm{M} = 3$; and (iii) the peak intensity decreases with an increasing number of molecules, as evidenced by the decreasing peak height from $N_\mathrm{M} = 1$ to $N_\mathrm{M} = 3$.

\subsection{Analysis of LFs for Molecules above a Silver Surface}

\subsubsection{Quantitative Analysis}

To understand these unusual phenomena, it is crucial to derive the analytical solutions of the LFs, which are governed by the quantum dynamics of the molecules. Based on the dynamical equation under Markov approximation in Eq.~(\ref{Eq:non_Hermitian_Schr_Eq}), we can derive the analytical solutions of the LFs in Eq.~(\ref{Eq:LF_11}) for the monomer, the dimer, and trimer systems as follows (see Appendix \ref{Appendix_Analytical_Solution}
for details):
\begin{align}
\nonumber
  &\mathrm{Monomer}: \\
\label{Eq:Analytical_Solution_LF3_1}
  &  \mathcal{L}^{(1)}(\omega_\mathrm{D}) = \frac{1}{ \left(\omega_\mathrm{D}-\omega_\mathrm{M} \right)^2+ \left(\frac{\Gamma}{2}\right)^2}, \\
\nonumber
  &\mathrm{Dimer}: \\
\label{Eq:Analytical_Solution_LF3_2}
  &  \mathcal{L}^{(2)}(\omega_\mathrm{D}) = \frac{2}{ \left(\omega_\mathrm{D}-\omega_\mathrm{M}-\mathrm{Re}\{\frac{V}{\hbar}\} \right)^2+\left(\frac{\Gamma}{2}-\mathrm{Im}\{\frac{V}{\hbar}\}\right)^2}, \\
\nonumber
  &\mathrm{Trimer}: \\
\nonumber
  &\mathcal{L}^{(3)}(\omega_\mathrm{D}) \\
  &\approx \frac{ 3 }{ \left(\omega_\mathrm{D}- \omega_\mathrm{M} -\mathrm{Re}\left\{\frac{V'+V_\mathrm{eff}}{2\hbar}\right\}  \right)^2 + \left(\frac{\Gamma}{2} - \mathrm{Im}\left\{\frac{V'+V_\mathrm{eff}}{2\hbar} \right\} \right)^2 }, 
\label{Eq:Analytical_Solution_LF3}
\end{align}
where the transition frequency $\omega_\mathrm{M}=\omega_1=\omega_2=\omega_3$ and the decay rates $\Gamma=\Gamma_1=\Gamma_2=\Gamma_3$ are the same for all molecules across all multichromophoric systems. The symbol $V$ represents the resonant dipole-dipole interaction between a pair of nearest-neighbor molecules, i.e., $V=V_{\mathrm{RDDI},12}=V_{\mathrm{RDDI},21}=V_{\mathrm{RDDI},23}=V_{\mathrm{RDDI},32}$ defined in Eq.~(\ref{Eq:RDDI}). The symbol $V'=V_{\mathrm{RDDI},13}=V_{\mathrm{RDDI},31}$ denotes the resonant dipole-dipole interaction between the first and the third molecule. The symbols $V_\mathrm{eff} = \sqrt{8V^2+V'^2}$ represents an effective interaction that combines contributions from both $V$ and $V'$. For the multichromophoric systems shown in Fig~\ref{Fig_2}, the values of these parameters are summarized in Table~\ref{tab:parameters}.
\begin{table}[h]
    \centering
    \begin{tabular}{c | c | c | c }
        \toprule
        Parameter & Value & Parameter & Value \\
         \hline
    $\hbar\omega_\mathrm{M}$ & 3.659 & $V'$ & $-0.586 - 0.038i$ \\
        $\hbar \Gamma$ & 0.148 & $V_\mathrm{eff}$ & $-13.140 - 0.180i$ \\
        $V$ & $-4.641 - 0.063i$ & &  \\
        \toprule
    \end{tabular}
    \caption{Parameters in Eqs.~(\ref{Eq:Analytical_Solution_LF3_1})-(\ref{Eq:Analytical_Solution_LF3}). $\hbar\omega_\mathrm{M}$ is in eV; $\hbar\Gamma$, $V$, $V'$, and $V_\mathrm{eff}$ are in meV. }
    \label{tab:parameters}
\end{table}

In Fig.~\ref{Fig_4}, we plot the analytical solutions for the LFs of all three systems using the yellow, green, and purple dashed lines, which are labeled as ``Ana.'' in the legend. The excellent agreement between the LFs obtained from the analytical solutions in Eqs.~(\ref{Eq:Analytical_Solution_LF3_1})-(\ref{Eq:Analytical_Solution_LF3}) and those from the full quantum dynamical simulations in Eqs.~(\ref{Eq:ODE_Equation}), (\ref{Eq:Two-Time_Correlation_App}), and (\ref{Eq:LF_11}) validate the accuracy of our analytical solutions. 

According to Eqs.~(\ref{Eq:Analytical_Solution_LF3_1})-(\ref{Eq:Analytical_Solution_LF3}), the peak positions for the monomer, the dimer, and the trimer systems can be found as $\omega_\mathrm{M}$, $\omega_\mathrm{M}+\Re{V}/\hbar$, and $\omega_\mathrm{M}+\Re{V'+V_\mathrm{eff}}/2\hbar$, respectively. Therefore, the redshift observed in the peak positions results from the negative real part of the resonant dipole-dipole interactions, e.g., $\Re{V}<0$. Furthermore, these redshift energies match the lowest eigenvalues of the real part of the non-Hermitian Hamiltonians in Eqs.~(\ref{Eq_H_Eff2}) and (\ref{Eq_H_Eff3}), e.g., the lowest eigenvalue of $\Re{\tilde{\mathbf{H}}^{(3)}}$ is $\Re{V'+V_\mathrm{eff}}/2$. 

As for the second phenomenon, the broadening of peak width with an increasing number of molecules can be attributed to the imaginary part of the resonant dipole-dipole interaction (non-Hermitian interactions). From the analytical solutions in Eqs.~(\ref{Eq:Analytical_Solution_LF3_1})-(\ref{Eq:Analytical_Solution_LF3}), we observe that the LFs for all three systems are Lorentzian. Accordingly, the full width at half maximum $W^{(N_\mathrm{M})}$ for the monomer, dimer, and trimer systems are listed below:
\begin{align*}
& \mathrm{Monomer:} & W^\mathrm{(1)} &=  \Gamma, \numberthis \label{Monmer_Width} \\
& \mathrm{Dimer:} &
W^\mathrm{(2)} &=  \Gamma-2\mathrm{Im}\{V/\hbar\}, \numberthis  \\
& \mathrm{Trimer:} & W^\mathrm{(3)} &=
\Gamma-\mathrm{Im}\{V'+V_\mathrm{eff}\}/\hbar, \numberthis \label{Trimer_Width}
\end{align*}   
Note that the broadening of the dimer and trimer systems depends on the imaginary part of the resonant dipole-dipole interactions, e.g., $V=V_\mathrm{RDDI,12}$, which is explicitly influenced by the molecular positions ($\mathbf{r}_1$ and $\mathbf{r}_2$), as shown in Eq.~(\ref{Eq:RDDI}). Moreover, the FWHM of the Lorentzian peak directly reflects the population decay rate in our system. For instance, using the coefficients from Eq.~(\ref{Eq:APP_Analytical_Solution_2}), the population decay of the dimer system is expressed as:
\begin{align}
P(t) = \sum_{\alpha=1}^2 \abs{C^{\mathrm{E}_\alpha,\{0\}}(t) }^2 = \exp{- \left( \Gamma-2\Im{\frac{V}{\hbar}} \right) t } .
\end{align}
Thus, the population decay rate is $\Gamma - 2\Im{V}/\hbar$, which matches the FWHM of the dimer system. Based on Eqs.~(\ref{Monmer_Width})-(\ref{Trimer_Width}), an increase in FWHM with more molecules indicates a higher population decay rate for the molecular aggregate, which is a characteristic of superradiance. Therefore, superradiance can be identified by observing the broadening of the Lorentzian peak's FWHM. 

Another key aspect that requires analysis is the enhancements of the population decay rates (which, in this work, is equivalent to the FWHM of the peak). The enhancement ratios for the monomer, dimer, and trimer are found to be $\approx 1:1.9:2.5$, which deviates from the $1:2:3$ ratio predicted by the Tavis-Cummings model\cite{Tavis1968} (or Dicke model\cite{Dicke1954}). This discrepancy arises because the Tavis-Cummings model assumes all molecules couple a single photonic mode and that the wavelength is much larger than the spatial extent of the molecular aggregate\cite{wang_2023}—conditions that are not fully met for molecules on a silver surface. However, when all molecules are close enough ($\mathbf{r}_3 = \mathbf{r}_2 = \mathbf{r}_1$), the enhancement ratio can indeed show an $N_\mathrm{M}$-fold dependence, as the imaginary part of all intermolecular resonant dipole-dipole interactions, $\Im\{V_\mathrm{RDDI,\alpha\beta}\}$, equals half of the spontaneous emission rate $\Gamma_\alpha/2$:
\begin{align}
\nonumber
    \left.-\frac{\Im\{V_\mathrm{RDDI,\alpha\beta}\}}{\hbar} \right|_{\alpha=\beta} &=  \frac{\omega_\alpha^2}{\hbar\epsilon_0 c^2} \pmb{\mu}_\alpha \cdot \Im \overline{\overline{\mathbf{G}}}(\mathbf{r}_\alpha, \mathbf{r}_\alpha, \omega_\alpha) \cdot \pmb{\mu}_\alpha \\
    &= \frac{\Gamma_\alpha}{2}. 
\end{align}
Consequently, the population decay rates for the monomer, dimer, and trimer are $\Gamma$, $\Gamma - 2\Im{V}/\hbar = 2\Gamma$, and $\Gamma - \Im{V' + V_\mathrm{eff}}/\hbar = 3\Gamma$, resulting in an enhancement ratio of exactly $1:2:3$. Therefore, when the intermolecular distance is sufficiently small, the superradiant behavior aligns with the predictions of the Tavis-Cummings model.

Turning to the third phenomenon, the peak intensity of LFs depends on two factors: the half width at half maximum (HWHM) and the numerator of the Lorentz functions. More specifically, according to Eqs.~(\ref{Eq:Analytical_Solution_LF3_1})-(\ref{Eq:Analytical_Solution_LF3}), the peak intensities $I^{(N_\mathrm{M})}$ of all three systems are given by:
\begin{align*}
    &\mathrm{Monomer}: & {I}^{(1)} &= \frac{1}{(W^{(1)})^2}, \qquad  \numberthis \label{Eq:Analytical_LF2_Height_1} \\
   &\mathrm{Dimer}:  & {I}^{(2)} &= \frac{2}{(W^{(2)})^2}, \qquad \numberthis \\
  & \mathrm{Trimer}:  & {I}^{(3)} &\approx \frac{ 3 }{ (W^{(2)})^2 }. \qquad \numberthis
\label{Eq:Analytical_LF2_Height}
\end{align*}
According to our previous discussion, the ratio of the denominator among the three systems is about $(W^{(1)})^2:(W^{(2)})^2:(W^{(3)})^2=1:3.6:6.3$. The ratio of the numerators in Eqs.~(\ref{Eq:Analytical_LF2_Height_1})-(\ref{Eq:Analytical_LF2_Height}) is close to $1:2:3$, which is consistent with the values obtained using Kasha's model for the square of the transition dipole matrix elements. Therefore, the ratio of peak heights for 
and the relative peak intensity can be estimated as $ I^{(1)}:I^{(2)}:I^{(3)} \approx 1:0.6:0.5$, which closely matches the observed ratio in Fig.~\ref{Fig_4}. Here, we demonstrate that this imaginary part of the resonant dipole-dipole interactions can also reduce the peak intensities of LFs in the emission power spectra of multichromophoric systems.

\subsubsection{Qualitative Analysis}

The redshifts in peak positions can be qualitatively explained using Kasha’s model\cite{KashaRawlsAshrafElBayoumi,hestand_2018},  which predicts the peak's redshifts  for the dimer and trimer as the lowest eigenvalues of the corresponding effective Hermitian Hamiltonians $\tilde{\mathbf{H}}^{(2)}_\mathrm{Kasha}$ and $\tilde{\mathbf{H}}^{(3)}_\mathrm{Kasha}$:
\begin{align}
\label{Eq_H_Ka_Eff2}
 &\tilde{\mathbf{H}}^{(2)}_\mathrm{Kasha} = \begin{bmatrix}
        0 & J_{12} \\
          J_{12} &  0  \\
    \end{bmatrix}, \\
 &\tilde{\mathbf{H}}^{(3)}_\mathrm{Kasha} = \begin{bmatrix}
           0  & J_{12} & J_{13} \\
        J_{12} &   0  & J_{23}\\
        J_{13} & J_{23} & 0 \\
    \end{bmatrix},
\label{Eq_H_Ka_Eff3}
\end{align}
where $J_{\alpha\beta}$ represents the dipole-dipole interaction potential between the $\alpha$ and $\beta$ molecules\cite{Wang2022_MC}:
\begin{align}
\label{Eq:J_Exciton_Ham}
   & J_{\alpha\beta} =   \frac{ \pmb{\mu}_\alpha \cdot \pmb{\mu}_\beta - 3(\pmb{\mu}_\alpha \cdot \mathbf{r}_{\alpha\beta} )(\pmb{\mu}_\beta \cdot \mathbf{r}_{\alpha\beta} )/\abs{\mathbf{r}_{\alpha\beta}}^2 }{ 4\pi\epsilon_0 \abs{\mathbf{r}_{\alpha\beta}}^3 },
\end{align}
where $\mathbf{r}_{\alpha\beta} = \mathbf{r}_\alpha - \mathbf{r}_\beta$ is the position vector from molecule $\beta$ to molecule $\alpha$. For the multichromophoric systems shown in Fig~\ref{Fig_2}, we have $J_{12}=J_{23}=-4.623$ meV and $J_{13}=-0.578$ meV. Since $J_{12}\approx \Re{V}$ and $J_{13}\approx \Re{V'}$, we can infer that the lowest eigenvalues of the Hermitian Hamiltonians $\tilde{\mathbf{H}}^{(2)}_\mathrm{Kasha}$ and $\tilde{\mathbf{H}}^{(3)}_\mathrm{Kasha}$ are nearly identical to those of the real parts of the non-Hermitian Hamiltonians $\Re{\tilde{\mathbf{H}}^{(2)}}$ and $\Re{\tilde{\mathbf{H}}^{(3)}}$ in Eqs.~(\ref{Eq_H_Eff2})-(\ref{Eq_H_Eff3}). Our results indicate that the real part of the non-Hermitian Hamiltonian can be effectively reduced to Kasha's model, which has been extensively applied in the study of photosynthesis and molecular aggregates. 

The phenomenon that the peak intensity decreases with the number of molecules can be understood as follows:

First, the area of emission power spectra corresponds to the photon energy (related to the number of photons). In our study, all initial states are 1-excitation states (delocalized exciton states), meaning the number of photons is the same across all three systems. If the surface does not significantly absorb photons, the areas of emission power spectra in the three systems are supposed to be nearly the same.

Second, the peak width corresponds to the population decay rate. In our study, we do not consider dissipation or coherence effects from other degrees of freedom, such as molecular vibrations. Therefore, the population decay only comes from the molecule-photon interactions. As the number of molecules increases, the delocalized exciton states coupled to photon modes lead to superradiance, meaning the population decay rate increases with the number of molecules. Quantitatively, the population decay rate, which is proportional to the number of molecules, can be understood via Dick’s $N$ scaling law\cite{Dicke1954}. Alternatively, the population decay rate can be also understood by our theoretical analysis: the population decay rate depends on the spontaneous emission rate and the imaginary part of distance-dependent dipole-dipole interactions (non-Hermitian effect), which also explains why the population decay does not follow Dick’s $N$ scaling law. For these two reasons, the peak width is approximately proportional to the number of molecules.

Third, if one agrees that the two statements (i) the areas of emission power spectra are supposed to be nearly the same, and (ii) the peak width is approximately proportional to the number of molecules, one can easily deduce that the peak height decreases with the number of molecules. 

In summary, our quantitative and qualitative analysis of the LFs for molecular aggregates weakly coupled to surface plasmon polaritons above a silver surface reveals three key findings: (i) The lowest eigenvalue of the real part of the non-Hermitian Hamiltonian in Eqs.~(\ref{Eq_H_Eff2})-(\ref{Eq_H_Eff3}) determines the peak redshifts. (ii) The imaginary part of the non-Hermitian Hamiltonian, including both the diagonal element $\Gamma_\mathrm{\alpha}$ and the off-diagonal element $V_\mathrm{RDDI,\alpha\beta}$, contributes to the broadening of peak widths. (iii) The imaginary part of the off-diagonal element enhances the population decay rate of molecular aggregates (i.e., superradiance), which is highly dependent on the molecular positions in complex dielectric environments.

\subsection{EEFs of Molecules above Silver Surface in the Far-Field Region}
\label{SubSec:EEF}

\begin{figure}[htbp] \includegraphics[width=0.5\textwidth]{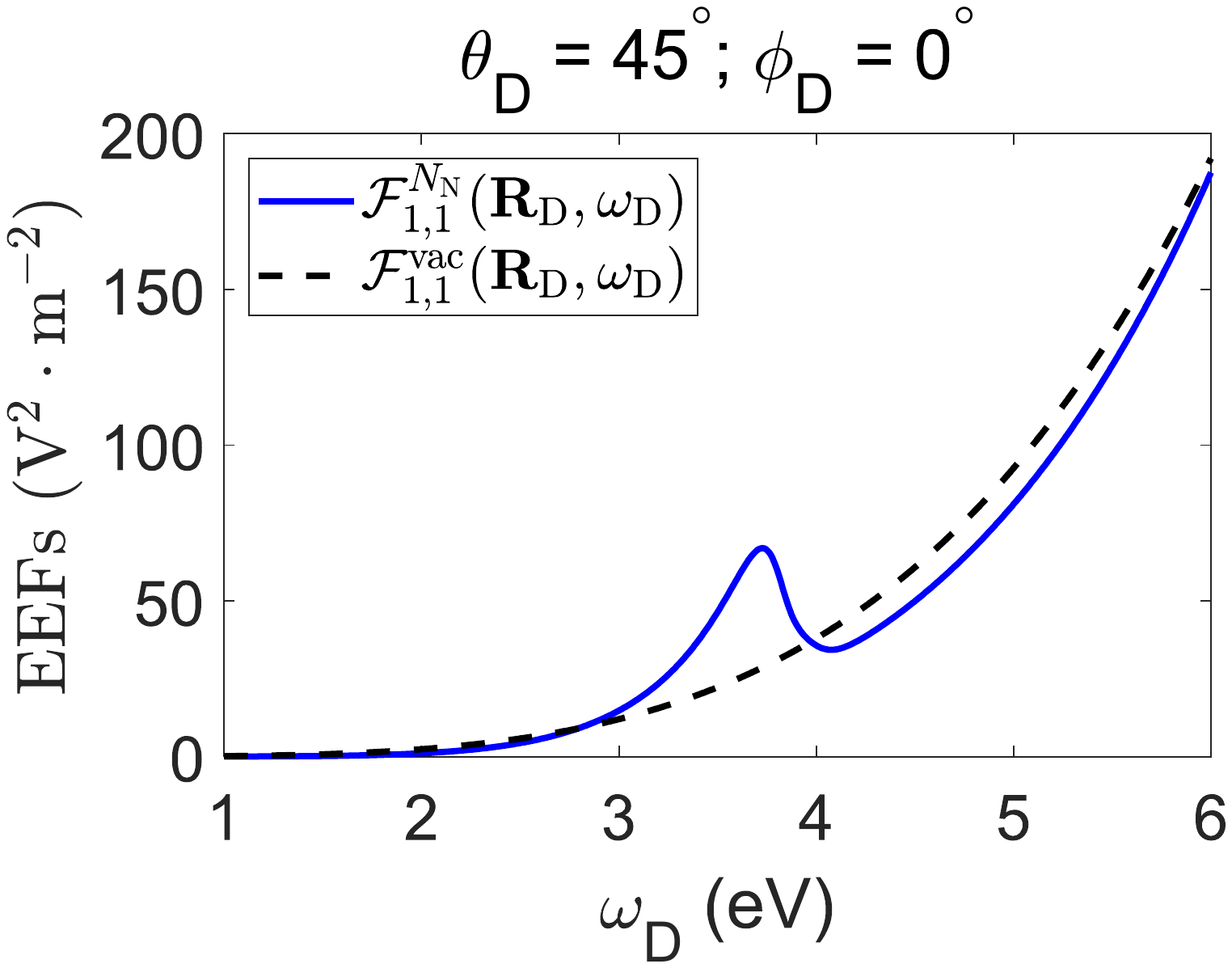}
\caption{Frequency dependence of the EEFs of molecular emitters oriented along the $x$-axis in the far-field region. The detector is positioned at the polar angle $\theta_\mathrm{D}=45^\circ$ and the azimuthal angle $\phi_\mathrm{D}=0^\circ$. The blue solid line represents the EEF of molecules above the silver surface, while the black dashed line indicates the EEF in vacuum. }
\label{Fig_5}
\end{figure}

In Fig~\ref{Fig_5}, we numerically demonstrate the frequency dependence of the specific EEF of molecules above a silver surface in the far-field region (recall Eq.~(\ref{Eq:EEF_11})) with a fixed polar angle $\theta_\mathrm{D}=45^\circ$ and azimuthal angle $\phi_\mathrm{D}=0^\circ$. To calculate the term $\overline{\overline{\mathbf{G}}}_\mathrm{far}(\mathbf{R}_\mathrm{D},\mathbf{r}_1,\omega_\mathrm{D}) \cdot {\pmb{\mu}}_1$ in Eq.~(\ref{Eq:EEF_11}) and generate Fig~\ref{Fig_5}, we set the magnitude of the molecular transition dipole moments $\abs{{\pmb{\mu}}_1}$ to 10 Debye, with transition dipoles oriented along the $x$-axis. Clearly, $\mathcal{F}^{(N_\mathrm{M})}_{1,1}(\mathbf{R}_\mathrm{D},\omega_\mathrm{D})$ increases with the detection frequency $\omega_\mathrm{D}$, showing a notable peak around 3.7 eV before continuing to rise steadily. The reason for this behavior can be divided into two parts: firstly, the steady increase is primarily due to the contribution of the free-space dyadic Green's function, as confirmed by the increase of $\mathcal{F}^\mathrm{vac}_{1,1}(\mathbf{R}_\mathrm{D},\omega_\mathrm{D})$ with frequency. Here, $\mathcal{F}^\mathrm{vac}_{1,1}(\mathbf{R}_\mathrm{D},\omega_\mathrm{D})$ is defined as $\mathcal{F}^{(N_\mathrm{M})}_{1,1}(\mathbf{R}_\mathrm{D},\omega_\mathrm{D})$ in Eq.~(\ref{Eq:EEF_11}), where the dyadic Green's function in Eq.~(\ref{Eq:Total_Greens_Func}) replaced by the free-space dyadic Green's function in Eq.~(\ref{Eq:Green_Func_Vac}). Secondly, the peak around 3.7 eV is attributed to the interference of the electric fields generated by the emitter dipole and its mirror dipole, as analyzed in our previous study\cite{lee_2021}.

In summary, our numerical result shows that the specific EEF of molecular aggregates placed above the silver surface has a significant impact on the observed emission power spectrum within the 3–4 eV photon frequency range. Therefore, when studying spectra within this frequency range, the influence of the EEF should be carefully considered. In our multichromophoric systems, since the peaks of the LFs are distributed within a narrow frequency range of 3.65 eV to 3.66 eV, it can be expected that the enhancement by EEF will not differ significantly across different peaks. Therefore, we do not provide an additional figure for the EPS.

\section{Conclusions}
\label{Sec:Conclusion}

Inspired by previous works\cite{wang2020theory,medina_2021,chuang_2024}, we have generalized the theory of molecular emission power spectra from one molecule to multichromophoric systems in the framework of MQED to investigate molecular spontaneous emission processes. This extension provides an alternative approach to characterizing light-matter hybrid states of multiple molecules in complex dielectric materials that are inhomogeneous, dispersive, and absorbing. Based on the total Hamiltonian of multichromophoric systems (Eqs.~(\ref{Eq:MQED_Molecules})-(\ref{Eq:Light-Matter_Interaction})), we present the full quantum dynamics within the truncated state subspace\cite{chuang_2024}, incorporating the counter-rotating effects. In the Heisenberg picture, we derive a general theory of the emission power spectra (Eq.~(\ref{Eq:EPS_Exact})) for multichromophoric systems, which is expressed as a sum of products of the exact EEFs (Eq.~(\ref{Eq:EEF_Exact})) and the exact LFs (Eq.~(\ref{Eq:Lineshape_Func})). This formula is not constrained by the truncated state vector. The exact LFs can be numerically estimated within this truncated state subspace, resulting in the expression provided in Eq.~(\ref{Eq:Two-Time_Correlation_App}), which is related to the quantum dynamics of the coefficients $C^{\mathrm{E}_{\alpha},\{0\}}(t')$. Based on Kramers-Kronig relations, the exact EEFs can be calculated by Eq.~(\ref{Eq:GEEF_Reduced}). Equations~(\ref{Eq:EPS_Exact}), (\ref{Eq:Two-Time_Correlation_App}), and (\ref{Eq:GEEF_Reduced}) are the main formula for calculating the emission power spectrum in complex dielectric environments, including nanophotonic structures where electric field profiles can vary significantly over short distances. For closely aligned molecular aggregates above a planar surface, we prove that the analytical EEFs can be approximated by the specific EEF in Eq.~(\ref{Eq:EEFs_for_Closely_Aligned}). Additionally, the corresponding specific LF in Eq.~(\ref{Eq:LF_11}) can be expressed as a sum of $\mathcal{L}^{(N_\mathrm{M})}_{\alpha,\beta}(\omega_\mathrm{D})$. Consequently, the EPS for planar systems is simplified to a single product of the specific EEFs (Eq.~(\ref{Eq:EEF_Exact})) and the specific LFs.

To demonstrate the polariton-induced collective effect on the emission power spectrum, we calculate the specific LF and the specific EEF for one to three molecules weakly coupled with surface plasmon polaritons above a silver surface. We summarize our main findings as follows. First, our study successfully captures the redshifts of peak positions in LF, which are related to the lowest eigenvalues of the real parts of the non-Hermitian Hamiltonians as described in Eqs.~(\ref{Eq_H_Eff2})-(\ref{Eq_H_Eff3}). Second, the broadening of peak widths results from the spontaneous emission rates and the imaginary part of the resonant dipole-dipole interactions (i.e., $\Gamma_\alpha$ and $V_\mathrm{RDDI,\alpha\beta}$). Moreover, without other degrees of freedom (e.g., molecular vibrations), the population decay rate of molecular aggregate corresponds to the superradiance rate, indicating that the superradiance rate depends on non-Hermitian interactions. Third, our study reveals that non-Hermitian interactions $\Gamma_\alpha$ and $V_\mathrm{RDDI,\alpha\beta}$ can be controlled through the molecules in a complex dielectric environment. Fourth, based on the qualitative and quantitative analyses, our study successfully explains why peak height decreases as $N_\mathrm{M}$ increases. Fifth, the far-field EEFs exhibit strong frequency dependence and significantly influence the emission power spectrum of molecules above a silver surface, particularly within the 3–4 eV photon frequency range. To sum up, through the analysis of the LF and EEF, we have demonstrated that the emission power spectra are able to characterize the quantum dynamics of molecular aggregates coupled with polaritons in a complex dielectric environment. We would like to emphasize that the general form of our theory can be used for molecules in the strong light-matter coupling regime, providing an alternative method to characterize light-matter hybrid states.

Although we have successfully explained the origin of the peak redshifts, decreased peak height, and broadened peak width in the emission power spectra in the weak light-matter coupling regime, this is not a whole story. First, we believe that an analysis of emission power spectra in the strong light-matter coupling regime is required for the field of polariton chemistry and quantum electrodynamic chemistry. Second, the current model Hamiltonian does not account for molecular vibrations or structural disorder, limiting its ability to capture vibrational substructures and broad spectral bandwidths observed in real molecules. However, this work can be viewed as a bridge between one molecule (Part I and Part II) and multiple molecules with vibrations and disorders (Future work). Third, we demonstrate the LFs (Eq.~(\ref{Eq:LF_11})) and EEFs (Eq.~(\ref{Eq:EEF_11})) for multiple molecules above a planar surface. The expressions of Eqs.~(\ref{Eq:LF_11}) and (\ref{Eq:EEF_11}) are derived under two approximations: the far-field approximation (i.e., ${ \left| \mathbf{R}_\mathrm{D}-\mathbf{r}_{1} \right| \gg \lambda_\mathrm{D} = 2\pi c/\omega_\mathrm{D} }$) and the assumption that the molecular aggregate’s length is much smaller than the axial radius between the molecule and the detector (i.e., $2d\ll \rho $). These approximations are specific to planar surfaces. When molecules are near nanophotonic structures, where electric field profiles can vary significantly over a few nanometers, Eqs.~(\ref{Eq:LF_11}) and (\ref{Eq:EEF_11}) may become invalid. Investigating the emission power spectra of multichromophoric systems in such nanophotonic structures also deserves future research. These limitations will be addressed in our future work. Macroscopic QED is a powerful tool for studying the hybrid states formed of molecules and photons, enabling control over fundamental physical and chemical processes such as electron transfer\cite{Semenov2019,Mandal2020,Phuc2020,Wellnitz2021,Wei2022,wei_2024} and energy transfer\cite{Green2020,Georgiou2021,Wang2021}. We anticipate that this study will serve as a foundation for further investigations into the emission power spectra and offer an alternative
approach to analyze the light-matter hybrid state of multichromophoric systems (or multiple emitters) coupled with polaritons in complicated nanostructures.

\begin{acknowledgments}
Wang thanks the National Center for High-performance Computing and the Core Facility for Networking Science, Technology, and advanced Computing for their high-performance computing resources, and Dr. Qian-Rui Huang and Yi-Ting Chuang for assisting with the calculations of quantum dynamics of multiple molecules in dielectric environments. Hsu thanks Academia Sinica (AS-CDA-111-M02), National Science and Technology Council (111-2113-M-001-027-MY4), and Physics Division, National Center for Theoretical Sciences (113-2124-M-002-003) for the financial support.
\end{acknowledgments}


\section*{data availability}
The data that support the findings of this study are available from the corresponding author upon reasonable request.

\begin{appendix}

\section{Derivation of Eq.~(\ref{Eq:E(+)_Integral_Form})}
\label{Appendix_Eq:E(+)_Integral_Form}
To derive Eq.~(\ref{Eq:E(+)_Integral_Form}), we begin with the total Hamiltonian $\hat{H}_\mathrm{T}$ composed of Eqs.~(\ref{Eq:MQED_Molecules})-(\ref{Eq:Light-Matter_Interaction}). In the Heisenberg picture, the time-dependent dynamics of an operator follows:
\begin{align}
    \frac{\partial }{\partial t} \hat{O}(t) = \frac{i}{\hbar}\left[ \hat{H}_\mathrm{T}(t) ,  \hat{O}(t) \right].
\end{align}

For the annihilation operator $\hat{\mathbf{f}}(\mathbf{r},\omega)$ in Eq.~(\ref{Eq:MQED_Polariton}), which commutes with the molecular Hamiltonian $\hat{H}_\mathrm{M}$, its equation of motion becomes
\begin{align}
    \frac{\partial }{\partial t} \hat{\mathbf{f}}(\mathbf{r},\omega,t) = \frac{i}{\hbar}\left[ \hat{H}_\mathrm{P}(t) + \hat{H}_\mathrm{M-P}(t) ,  \hat{\mathbf{f}}(\mathbf{r},\omega,t) \right].   
\label{Eq:Append_EQofMotion}
\end{align}

Substituting Eqs.~(\ref{Eq:MQED_Polariton}) and (\ref{Eq:Light-Matter_Interaction}) into Eq.~(\ref{Eq:Append_EQofMotion}), and using the commutation relation $\left[ \hat{\mathrm{f}}_i(\mathbf{r},\omega) , \hat{\mathrm{f}}^\dagger_j(\mathbf{r}',\omega') \right] = \delta_{ij} \delta(\mathbf{r}-\mathbf{r}')\delta(\omega-\omega') $, we obtain
\begin{align}
\label{Eq:App_f}
    &\frac{\partial }{\partial t} \hat{\mathbf{f}}(\mathbf{r},\omega,t) =  -i\omega  \hat{\mathbf{f}}(\mathbf{r},\omega,t) +\frac{i}{\hbar}  \sum_{\alpha=1}^{N_\mathrm{M}} \hat{\pmb{\mu}}_\alpha(t) \cdot \overline{\overline{\mathbf{g}}}^*(\mathbf{r}_\alpha,\mathbf{r},\omega) .
\end{align}

Similarly, the equation of motion for the creation operator $\hat{\mathbf{f}}^\dagger(\mathbf{r},\omega)$ is given by:
\begin{align}
\label{Eq:App_f^dagger}
    &\frac{\partial }{\partial t} \hat{\mathbf{f}}^\dagger(\mathbf{r},\omega,t) =  i\omega  \hat{\mathbf{f}}^\dagger(\mathbf{r},\omega,t) -\frac{i}{\hbar}  \sum_{\alpha=1}^{N_\mathrm{M}} \hat{\pmb{\mu}}_\alpha(t) \cdot \overline{\overline{\mathbf{g}}}(\mathbf{r}_\alpha,\mathbf{r},\omega).
\end{align}

By formally integrating Eqs.~(\ref{Eq:App_f}) and (\ref{Eq:App_f^dagger}), we obtain
\begin{align}
\nonumber
    &\, \hat{\mathbf{f}}(\mathbf{r},\omega,t) \\
\label{Eq:App_f_Integral}
    &= \hat{\mathbf{f}}(\mathbf{r},\omega,0)e^{-i\omega t} + \frac{i}{\hbar}  \sum_{\alpha=1}^{N_\mathrm{M}} \int_0^t dt' 
    e^{-i\omega(t-t')} \hat{\pmb{\mu}}_\alpha(t') \cdot \overline{\overline{\mathbf{g}}}^*(\mathbf{r}_\alpha,\mathbf{r},\omega), \\
\nonumber
    &\, \hat{\mathbf{f}}^\dagger(\mathbf{r},\omega,t) \\
    &= \hat{\mathbf{f}}^\dagger(\mathbf{r},\omega,0)e^{i\omega t} - \frac{i}{\hbar}  \sum_{\alpha=1}^{N_\mathrm{M}} \int_0^t dt' 
    e^{i\omega(t-t')} \hat{\pmb{\mu}}_\alpha(t') \cdot \overline{\overline{\mathbf{g}}}(\mathbf{r}_\alpha,\mathbf{r},\omega).
\end{align}

\begin{widetext}
Substituting Eq.~(\ref{Eq:App_f_Integral}) into Eq.~(\ref{Eq:E(+)}), we can express Eq.~(\ref{Eq:E(+)}) in the Heisenberg picture as:
\begin{align}
    \hat{\mathbf{E}}^{(+)}(\mathbf{R}_\mathrm{D},t_1) = & \int \mathrm{d}\mathbf{r}   \int_0^\infty d\omega \,  \overline{\overline{\mathbf{g}}}(\mathbf{R}_\mathrm{D},\mathbf{r},\omega) \cdot \hat{\mathbf{f}}(\mathbf{r},\omega,t_1) \\
    =& \hat{\mathbf{E}}^{(+)}_\mathrm{free}(\mathbf{R}_\mathrm{D},t_1)  +  \frac{i}{\hbar}  \sum_{\alpha=1}^{N_\mathrm{M}} \int_0^{t_1} dt' 
   \int \mathrm{d}\mathbf{r}   \int_0^\infty d\omega \,  \overline{\overline{\mathbf{g}}}(\mathbf{R}_\mathrm{D},\mathbf{r},\omega) \cdot \left\{  \hat{\pmb{\mu}}_\alpha(t') \cdot \overline{\overline{\mathbf{g}}}^*(\mathbf{r}_\alpha,\mathbf{r},\omega)   e^{-i\omega(t_1-t')} \right\}, 
\label{Eq:App_E(+)_Derivation_Inter}
\end{align}
\end{widetext}
where
\begin{align}
      \hat{\mathbf{E}}^{(+)}_\mathrm{free}(\mathbf{R}_\mathrm{D},t_1) &= \int \mathrm{d}\mathbf{r}   \int_0^\infty d\omega \,  e^{-i\omega t_1}\, \overline{\overline{\mathbf{g}}}(\mathbf{R}_\mathrm{D},\mathbf{r},\omega) \cdot\hat{\mathbf{f}}(\mathbf{r},\omega,0).
\end{align}

The second term in Eq.~(\ref{Eq:App_E(+)_Derivation_Inter}) can be further simplified by using the relationship of the auxiliary tensor functions\cite{Dung1998},
\begin{align}
\sum_{i=1}^3 \int d\mathbf{r} \; \overline{\overline{{g}}}_{ji}(\mathbf{R}_\mathrm{D},\mathbf{r},\omega)\overline{\overline{{g}}}_{ki}^*(\mathbf{r}_\alpha,\mathbf{r},\omega) =\frac{\hbar}{\pi\epsilon_0}\frac{\omega^2}{c^2} \mathrm{Im} \overline{\overline{G}}_{jk}(\mathbf{R}_\mathrm{D},\mathbf{r}_\mathrm{\alpha},\omega),  
\label{Eq:Green_function_Normalization}
\end{align}
where $\overline{\overline{{g}}}_{ji}(\mathbf{R}_\mathrm{D},\mathbf{r},\omega)$ and $\overline{\overline{{g}}}_{ki}^*(\mathbf{r}_\alpha,\mathbf{r},\omega)$ are the matrix elements of the auxiliary tensor functions $\overline{\overline{\mathbf{g}}}(\mathbf{R}_\mathrm{D},\mathbf{r},\omega)$ and $\overline{\overline{\mathbf{g}}}^*(\mathbf{r}_\alpha,\mathbf{r},\omega)$, respectively. Similarly, $\overline{\overline{G}}_{jk}(\mathbf{R}_\mathrm{D},\mathbf{r}_\alpha,\omega)$ is the matrix element of the dyadic Green's function $\overline{\overline{\mathbf{G}}}(\mathbf{R}_\mathrm{D},\mathbf{r}_\alpha,\omega)$. Substituting Eq.~(\ref{Eq:Green_function_Normalization}) into Eq.~(\ref{Eq:App_E(+)_Derivation_Inter}), we obtain Eq.~(\ref{Eq:E(+)_Integral_Form}) in the main text:
\begin{align}
\nonumber
  &\hat{\mathbf{E}}^{(+)}(\mathbf{R}_\mathrm{D},t_1)
   \\&=\hat{\mathbf{E}}^{(+)}_\mathrm{free}(\mathbf{R}_\mathrm{D},t_1)  + i \sum_{\alpha=1}^{N_\mathrm{M}} \int_0^{t_1} dt' \;  \mathbf{K}_\alpha (\mathbf{R}_\mathrm{D},t_1-t') 
   \hat{\bar{\mu}}_\alpha(t') ,
\end{align}
where
\begin{align}
      \mathbf{K}_\alpha (\mathbf{R}_\mathrm{D},\tau) &\equiv  \int_0^\infty d\omega \, \frac{\omega^2}{\pi\epsilon_0 c^2}  \mathrm{Im}\overline{\overline{\mathbf{G}}}(\mathbf{R}_\mathrm{D},\mathbf{r}_\alpha,\omega) \cdot {\pmb{\mu}}_\alpha e^{-i\omega\tau} .
\end{align}

\section{Derivation of Eq.~(\ref{Eq:Correlation_Express_2})}
\label{Appendix_Eq:Correlation_Express_2}

Since $\hat{\mathbf{E}}^{(-)}(\mathbf{R}_\mathrm{D},t_2)$ is the Hermitian conjugate of $\hat{\mathbf{E}}^{(+)}(\mathbf{R}_\mathrm{D},t_2)$ in Eq.~(\ref{Eq:App_E(+)_Derivation_Inter}), it can be expressed as
\begin{align}
\label{Eq:App_E(-)}
\nonumber
    &\hat{\mathbf{E}}^{(-)}(\mathbf{R}_\mathrm{D},t_2) \\ &= \hat{\mathbf{E}}^{(-)}_\mathrm{free}(\mathbf{R}_\mathrm{D},t_2)  - i \sum_{\beta=1}^{N_\mathrm{M}} \int_0^{t_2} dt'' \;  \mathbf{K}^*_\beta (\mathbf{R}_\mathrm{D},t_2-t'') \hat{\bar{\mu}}_\beta(t''),
\end{align}
with 
\begin{align} 
    &\hat{\mathbf{E}}^{(-)}_\mathrm{free}(\mathbf{R}_\mathrm{D},t_2) = \int \mathrm{d}\mathbf{r}   \int_0^\infty d\omega \,  e^{i\omega t_2}\, \overline{\overline{\mathbf{g}}}^*(\mathbf{R}_\mathrm{D},\mathbf{r},\omega) \cdot\hat{\mathbf{f}}^\dagger(\mathbf{r},\omega,0), \\
    &\mathbf{K}^*_\beta (\mathbf{R}_\mathrm{D},t_2-t'') = \int_0^\infty d\omega \, \frac{\omega^2}{\pi\epsilon_0 c^2}  \mathrm{Im}\overline{\overline{\mathbf{G}}}(\mathbf{R}_\mathrm{D},\mathbf{r}_\beta,\omega) \cdot {\pmb{\mu}}_\beta e^{i\omega(t_2-t'')}. 
\end{align}

Substituting Eqs.~(\ref{Eq:App_E(+)_Derivation_Inter}) and (\ref{Eq:App_E(-)}) into Eq.~(\ref{Eq:Time-Correlation_Def}), we can express the two-time correlation function as
\begin{align}
\nonumber
    &{C}(\mathbf{R}_\mathrm{D},t_1,t_2) \\
\nonumber
   &= \left< \hat{\mathbf{E}}^{(-)}_\mathrm{free}(\mathbf{R}_\mathrm{D},t_2) \cdot \hat{\mathbf{E}}^{(+)}_\mathrm{free}(\mathbf{R}_\mathrm{D},t_1) \right> \\
\nonumber
   &\quad + i\left< \sum_{\alpha=1}^{N_\mathrm{M}} \int_0^{t_1} dt' \; \hat{\mathbf{E}}^{(-)}_\mathrm{free}(\mathbf{R}_\mathrm{D},t_2) \cdot   \mathbf{K}_\alpha (\mathbf{R}_\mathrm{D},t_1-t') 
   \hat{\bar{\mu}}_\alpha(t') \right>  \\
\nonumber
   & \quad - i \left< \sum_{\beta=1}^{N_\mathrm{M}} \int_0^{t_2} dt'' \;  \mathbf{K}^*_\beta (\mathbf{R}_\mathrm{D},t_2-t'') \cdot \hat{\mathbf{E}}^{(+)}_\mathrm{free}(\mathbf{R}_\mathrm{D},t_1) \hat{\bar{\mu}}_\beta(t'')   \right> \\
\nonumber
   &\quad +  \sum_{\alpha,\beta=1}^{N_\mathrm{M}} \int_0^{t_1} dt' \int_0^{t_2} dt'' \; \mathbf{K}^*_\beta (\mathbf{R}_\mathrm{D},t_2-t'') \cdot    \mathbf{K}_\alpha (\mathbf{R}_\mathrm{D},t_1-t') \\
   &\qquad \times \left<   \hat{\bar{\mu}}_\beta(t'') 
   \hat{\bar{\mu}}_\alpha(t') \right>.
\label{Eq:Correlation_Express_1}
\end{align} 

In the Heisenberg picture, to estimate the expectation value in Eq.~(\ref{Eq:Correlation_Express_1}), we assume that the state vector at $t=0$ can be represented as the tensor product of the molecular electronic state $\ket{\psi_\mathrm{M}(0)}$ and the polariton state $\ket{\psi_\mathrm{P}(0)}$, i.e., $\ket{\Psi(0)}=\ket{\psi_\mathrm{M}(0)}\ket{\psi_\mathrm{P}(0)}$. Furthermore, if the polariton state is initially in the vacuum state $\ket{\psi_\mathrm{P}(0)}=\ket{\left\{0\right\}}$, the expectation value of the first term in Eq.~(\ref{Eq:Correlation_Express_1}) becomes
\begin{align}
\nonumber  
    &\left< {\Psi(0)} \right|  \hat{\mathbf{E}}^{(-)}_\mathrm{free}(\mathbf{R}_\mathrm{D},t_2) \cdot \hat{\mathbf{E}}^{(+)}_\mathrm{free}(\mathbf{R}_\mathrm{D},t_1) \left| {\Psi(0)}\right> \\ 
\nonumber
&\propto  \sum_{i,j,k=1}^3 \bra{\left\{0\right\}}\overline{\overline{{g}}}^*_{ij}(\mathbf{R}_\mathrm{D},\mathbf{r},\omega) \hat{{f}}_j^\dagger(\mathbf{r},\omega,0) \\
    & \qquad \qquad \times \overline{\overline{{g}}}_{ik}(\mathbf{R}_\mathrm{D},\mathbf{r},\omega) \hat{{f}}_k(\mathbf{r},\omega,0) \ket{\left\{0\right\}} = 0.
\label{Eq:Append_First_Term_Cor}
\end{align}

Equation (\ref{Eq:Append_First_Term_Cor}) equals zero since the annihilation operator acting on the vacuum state yields zero, i.e., $\hat{{f}}_k(\mathbf{r},\omega,0) \ket{\left\{0\right\}}=0$. Similarly, the second and third terms in Eq.~(\ref{Eq:Correlation_Express_1}) are also zero. Therefore, Eq.~(\ref{Eq:Correlation_Express_1}) reduces to Eq.~(\ref{Eq:Correlation_Express_2}) in the main text:
\begin{align}
\nonumber
    {C}(\mathbf{R}_\mathrm{D},t_1,t_2) = & \sum_{\alpha,\beta=1}^{N_\mathrm{M}} \int_0^{t_1} dt' \int_0^{t_2} dt'' \;  \left<   \hat{\bar{\mu}}_\beta(t'') 
   \hat{\bar{\mu}}_\alpha(t') \right> \\
   &  \times \mathbf{K}^*_\beta (\mathbf{R}_\mathrm{D},t_2-t'') \cdot    \mathbf{K}_\alpha (\mathbf{R}_\mathrm{D},t_1-t') .
\end{align} 

\section{Derivation of Eq.~(\ref{Eq:EPS_Exact})}
\label{Appendix_EPS_EXact}
To derive Eq.~(\ref{Eq:EPS_Exact}), we start by substituting Eq.~(\ref{Eq:Correlation_Express_2}) into Eq.~(\ref{Eq:Definiation_Power_Spectrum_Formal}) and then obtain
\begin{align}
\nonumber
    &\mathcal{S}^{(N_\mathrm{M})}(\mathbf{R}_\mathrm{D},\omega_\mathrm{D}) =   \sum_{\alpha,\beta=1}^{N_\mathrm{M}}  \int_0^\infty dt_2 \int_0^\infty dt_1 \int_0^{t_1} dt' \int_0^{t_2} dt'' \\
    & \times e^{-i\omega_\mathrm{D}(t_2-t_1)}  \mathbf{K}^*_\beta (\mathbf{R}_\mathrm{D},t_2-t'') \cdot    \mathbf{K}_\alpha (\mathbf{R}_\mathrm{D},t_1-t')  \left< \hat{\bar{\mu}}_\beta(t'') 
   \hat{\bar{\mu}}_\alpha(t')  \right> , \\
\nonumber
 &   = \sum_{\alpha,\beta=1}^{N_\mathrm{M}}  \int_0^\infty dt'' \int_0^{\infty} dt'  \int_{t'}^\infty dt_1 \int_{t''}^\infty dt_2 \; e^{-i\omega_\mathrm{D}(t_2-t_1)} \\
   & \quad \times \mathbf{K}^*_\beta (\mathbf{R}_\mathrm{D},t_2-t'') \cdot    \mathbf{K}_\alpha (\mathbf{R}_\mathrm{D},t_1-t')  \left<  \hat{\bar{\mu}}_\beta(t'') 
   \hat{\bar{\mu}}_\alpha(t')  \right>.
\label{Eq:App_Order_Integral}
\end{align}

Note that we change the order of the integration in Eq.~(\ref{Eq:App_Order_Integral}). Next, we define the time variables $\tau_1 = t_1-t'$ and $\tau_2 = t_2-t''$, allowing us to derive Eq.~(\ref{Eq:EPS_Exact}) in the main text:
\begin{align}
\nonumber
     \mathcal{S}^{(N_\mathrm{M})}(\mathbf{R}_\mathrm{D},\omega_\mathrm{D})   =&\sum_{\alpha,\beta=1}^{N_\mathrm{M}} \mathcal{F}_{\alpha,\beta}(\mathbf{R}_\mathrm{D},\omega_\mathrm{D}) \mathcal{L}_{\alpha,\beta}(\omega_\mathrm{D}),
\end{align}
with
\begin{align}
\nonumber
    &\mathcal{F}^{(N_\mathrm{M})}_{\alpha,\beta}(\mathbf{R}_\mathrm{D},\omega_\mathrm{D})=\int_{0}^\infty d\tau_1 \int_{0}^\infty d\tau_2 \; e^{-i\omega_\mathrm{D}(\tau_2-\tau_1)} \\
    &\qquad \qquad \qquad \quad \times \mathbf{K}^*_\beta (\mathbf{R}_\mathrm{D},\tau_2) \cdot    \mathbf{K}_\alpha (\mathbf{R}_\mathrm{D},\tau_1) , \\
     &\mathcal{L}^{(N_\mathrm{M})}_{\alpha,\beta}(\omega_\mathrm{D}) =   \int_0^\infty dt'' \int_0^{\infty} dt' \; e^{-i\omega_\mathrm{D}(t''-t')}  
 \left< \hat{\bar{\mu}}_\beta(t'') 
   \hat{\bar{\mu}}_\alpha(t') \right>  . 
\end{align}

\section{Derivation of Eq.~(\ref{Eq:GEEF_Reduced})}
\label{Appendix_Eq:GEEF_Reduced}
The Laplace transformation of $\mathbf{K}_\alpha (\mathbf{R}_\mathrm{D},\tau_1)$ in Eq.~(\ref{Eq:GEEF_Alternative_Form}) can be estimated using the Sokhotski-Plemelj theorem,
\begin{align}
\nonumber
    &\mathfrak{L}_{\tau_1}\left\{ \mathbf{K}_\alpha (\mathbf{R}_\mathrm{D},\tau_1) \right\}(-i\omega_\mathrm{D})\\
    &=  \int_0^\infty d\omega \, \int_0^\infty d\tau_1 \; e^{i(\omega_\mathrm{D}-\omega) \tau_1} \frac{\omega^2}{\pi\epsilon_0 c^2}  \mathrm{Im}\overline{\overline{\mathbf{G}}}(\mathbf{R}_\mathrm{D},\mathbf{r}_\alpha,\omega) \cdot {\boldsymbol{\mu}}_\alpha  \\
\nonumber
    &= \lim_{\zeta\rightarrow 0^+} \int_0^\infty d\omega \, \int_0^\infty d\tau_1 \; e^{i(\omega_\mathrm{D}-\omega+i\zeta) \tau_1} \frac{\omega^2}{\pi\epsilon_0 c^2} \\
    &\quad \times \mathrm{Im}\overline{\overline{\mathbf{G}}}(\mathbf{R}_\mathrm{D},\mathbf{r}_\alpha,\omega) \cdot {\boldsymbol{\mu}}_\alpha  \\
    &= \lim_{\zeta\rightarrow 0^+} \int_0^\infty d\omega \,  \frac{1}{i(\omega-\omega_\mathrm{D}-i\zeta)} \frac{\omega^2}{\pi\epsilon_0 c^2}  \mathrm{Im}\overline{\overline{\mathbf{G}}}(\mathbf{R}_\mathrm{D},\mathbf{r}_\alpha,\omega) \cdot {\boldsymbol{\mu}}_\alpha \\
\nonumber
    &=  \frac{\omega_\mathrm{D}^2}{\epsilon_0 c^2}  \mathrm{Im}\overline{\overline{\mathbf{G}}}(\mathbf{R}_\mathrm{D},\mathbf{r}_\alpha,\omega_\mathrm{D}) \cdot {\boldsymbol{\mu}}_\alpha \\
    & \quad - \frac{i}{\pi\epsilon_0 c^2} \mathcal{P}   \int_0^\infty d\omega \, \frac{\omega^2}{\omega-\omega_\mathrm{D} }\mathrm{Im}\overline{\overline{\mathbf{G}}}(\mathbf{R}_\mathrm{D},\mathbf{r}_\alpha,\omega) \cdot {\boldsymbol{\mu}}_\alpha,
\label{Eq:App_Laplace_K}
\end{align}
where the symbol $\mathcal{P}$ denotes the Cauchy principal value. To evaluate this Cauchy principle value, we assume that the contribution from region $(-\infty,0)$ can be neglected,
\begin{align}
    \nonumber
    &\mathcal{P}   \int_0^\infty d\omega \, \frac{\omega^2}{\omega-\omega_\mathrm{D} }\mathrm{Im}\overline{\overline{\mathbf{G}}}(\mathbf{R}_\mathrm{D},\mathbf{r}_\alpha,\omega) \cdot {\boldsymbol{\mu}}_\alpha \\
    &\approx \mathcal{P}  \int_{-\infty}^\infty d\omega \, \frac{\omega^2}{\omega-\omega_\mathrm{D} }\mathrm{Im}\overline{\overline{\mathbf{G}}}(\mathbf{R}_\mathrm{D},\mathbf{r}_\alpha,\omega) \cdot {\boldsymbol{\mu}}_\alpha \\
    &=
\pi\omega_\mathrm{D}^2\mathrm{Re}\left[\overline{\overline{\mathbf{G}}}(\mathbf{R}_\mathrm{D},\mathbf{r}_\alpha,\omega_\mathrm{D})\right]\cdot\boldsymbol{\mu}_\alpha. 
\label{Eq:App_ReG}
\end{align}

Here, we apply the Kramers-Kronig relation to derive Eq.~(\ref{Eq:App_ReG}). Substituting Eq.~(\ref{Eq:App_ReG}) into Eq.~(\ref{Eq:App_Laplace_K}), we obtain
\begin{align}
\nonumber
     &\mathfrak{L}_{\tau_1}\left\{ \mathbf{K}_\alpha (\mathbf{R}_\mathrm{D},\tau_1) \right\}(-i\omega_\mathrm{D})\\
     &\simeq \frac{\omega_\mathrm{D}^2}{\epsilon_0 c^2}\left\{ \mathrm{Im}\overline{\overline{\mathbf{G}}}(\mathbf{R}_\mathrm{D},\mathbf{r}_\alpha,\omega_\mathrm{D}) -i \mathrm{Re} \overline{\overline{\mathbf{G}}}(\mathbf{R}_\mathrm{D},\mathbf{r}_\alpha,\omega_\mathrm{D})  \right\} \cdot {\boldsymbol{\mu}}_\alpha  \\
    &=
    \frac{\omega_\mathrm{D}^2}{i\epsilon_0 c^2}\overline{\overline{\mathbf{G}}}(\mathbf{R}_\mathrm{D},\mathbf{r}_\alpha,\omega_\mathrm{D}) \cdot {\boldsymbol{\mu}}_\alpha .
    \label{Eq:bF}
\end{align}

Similarly, the Laplace transformation of $\mathbf{K}^*_\beta (\mathbf{R}_\mathrm{D},\tau_2)$  in Eq.~(\ref{Eq:GEEF_Alternative_Form}) becomes
\begin{align}
    \mathfrak{L}_{\tau_2}\left\{ \mathbf{K}^*_\beta (\mathbf{R}_\mathrm{D},\tau_2) \right\}(i\omega_\mathrm{D})  \simeq \frac{\omega_\mathrm{D}^2}{-i\epsilon_0 c^2} {\boldsymbol{\mu}}_\beta \cdot \overline{\overline{\mathbf{G}}}^\dagger(\mathbf{R}_\mathrm{D},\mathbf{r}_\beta,\omega_\mathrm{D}) .
\label{Eq:bF_star}
\end{align}

By substituting Eqs.~(\ref{Eq:bF}) and (\ref{Eq:bF_star}) into Eq.~(\ref{Eq:GEEF_Alternative_Form}), we obtain Eq.~(\ref{Eq:GEEF_Reduced}) in the main text:
\begin{align}
\nonumber
   & \mathcal{F}^{(N_\mathrm{M})}_{\alpha,\beta}(\mathbf{R}_\mathrm{D},\omega_\mathrm{D}) \\
    &\simeq \left( \frac{\omega_\mathrm{D}^2}{\epsilon_0 c^2} \right)^2 {\boldsymbol{\mu}}_\alpha \cdot \overline{\overline{\mathbf{G}}}^\dagger(\mathbf{R}_\mathrm{D},\mathbf{r}_\alpha,\omega_\mathrm{D})  \cdot   \overline{\overline{\mathbf{G}}}(\mathbf{R}_\mathrm{D},\mathbf{r}_\beta,\omega_\mathrm{D}) \cdot {\boldsymbol{\mu}}_\beta .
\end{align}

\section{Derivation of Eq.~(\ref{Eq:Two-Time_Correlation_App})}
\label{Appendix_Time_Correlation_Functino_Dipole}
Substituting the operator $\hat{\bar{\mu}}_{\alpha(\beta)}$ into Eq.~(\ref{Eq:Two_time_Schrodinger}), the time correlation function become
\begin{align}
\nonumber
 &\left< \hat{\bar{\mu}}_\beta(t'') 
   \hat{\bar{\mu}}_\alpha(t') \right>   \\
   &= \left< \Psi(t'') \right|   \left(\hat{\sigma}_{\beta}^{(-)}+ \hat{\sigma}_{\beta}^{(+)}\right)  e^{i\hat{H}_\mathrm{T}(t'-t'')/\hbar}
    \left(\hat{\sigma}_{\alpha}^{(-)}+ \hat{\sigma}_{\alpha}^{(+)}\right) \left| \Psi(t') \right>.
\label{Eq:Append_two_time_dipole}
\end{align}

To estimate Eq.~(\ref{Eq:Append_two_time_dipole}), we first calculate
\begin{align}
\label{Eq:Append_Sigma_ket_pm}
    \left( \hat{\sigma}_\gamma^{(+)} + \hat{\sigma}_\gamma^{(-)} \right) \ket{\Psi(t')} =  \hat{\sigma}_\gamma^{(+)} \ket{\Psi(t')} + \hat{\sigma}_\gamma^{(-)} \ket{\Psi(t')},
\end{align}
where $\gamma = \alpha$ or $\beta$, and
\begin{align}
\nonumber
    \hat{\sigma}_\gamma^{(-)} \ket{\Psi(t')}  =&  C^{\mathrm{E}_\gamma,\{0\}}(t') e^{-i W^{\mathrm{E}_\gamma,\{0\}} t'}\ket{\mathrm{G}} \ket{\left\{0\right\}}  +\sum_{\alpha=1}^{N_\mathrm{M}} \sum_{\beta>\alpha}^{N_\mathrm{M}} \sum_{k=1}^3 \\
\nonumber
   & \int d \mathbf{r} \int_0^{\infty} d\omega \, C^{\mathrm{E}_{\alpha \beta},\left\{1_k\right\}}(\mathbf{r}, \omega, t') e^{-i W^{\mathrm{E}_{\alpha \beta},\{1\}}(\omega) t'} \\
   &\times \hat{\sigma}_\gamma^{(-)} \ket{ \mathrm{E}_{\alpha \beta}} \ket{\left\{1_k(\mathbf{r}, \omega)\right\} }.
 \label{Eq:Apend_Sigma_ket}
\end{align}

Since $\left[\hat{\sigma}_\gamma^{(-)}, \hat{\sigma}_\alpha^{(+)} \right] = \delta_{\alpha,\gamma} \left( \ket{\mathrm{g}_\gamma }\bra{ \mathrm{g}_\gamma } - \ket{\mathrm{e}_\gamma }\bra{ \mathrm{e}_\gamma } \right) \equiv -\delta_{\alpha,\gamma} \hat{\sigma}^\mathrm{z}_\gamma  $, $\left[\hat{\sigma}_\gamma^\mathrm{z}, \hat{\sigma}_\beta^{(+)} \right] = -2\delta_{\beta,\gamma} \hat{\sigma}^\mathrm{(+)}_\gamma $, and $\hat{\sigma}^\mathrm{z}_\gamma \ket{\mathrm{G}} = -\ket{\mathrm{G}}$, the operator $ \hat{\sigma}_\gamma^{(-)}$ acting on the state $\ket{ \mathrm{E}_{\alpha \beta}}$ in Eq.~(\ref{Eq:Apend_Sigma_ket}) can be evaluated as
\begin{align}
    \hat{\sigma}_\gamma^{(-)} 
 \ket{ \mathrm{E}_{\alpha \beta}} &=  \hat{\sigma}_\gamma^{(-)} \hat{\sigma}_\alpha^{(+)} \hat{\sigma}_\beta^{(+)} \ket{\mathrm{G}}  \\
 &=\delta_{\beta,\gamma} \ket{\mathrm{E}_\alpha} + \delta_{\alpha,\gamma} \ket{\mathrm{E}_\beta} + 2\delta_{\alpha,\gamma}\delta_{\beta,\gamma} \ket{\mathrm{E}_\gamma} .
\label{Eq:Append_Sigma_g_E}
\end{align}

Substituting Eq.~(\ref{Eq:Append_Sigma_g_E}) into Eq.~(\ref{Eq:Apend_Sigma_ket}) and due to $\sum_{\alpha=1}^{N_\mathrm{M}} \sum_{\beta>\alpha}^{N_\mathrm{M}} 
\delta_{\alpha,\gamma}\delta_{\beta,\gamma} \ket{\mathrm{E}_\gamma} = 0 $, we obtain:
\begin{align}
\nonumber
    \hat{\sigma}_\gamma^{(-)} \ket{\Psi(t')}  
    =& C^{\mathrm{E}_\gamma,\{0\}}(t') e^{-i W^{\mathrm{E}_\gamma,\{0\}} t'}\ket{\mathrm{G}} \ket{\left\{0\right\}}  \\
\nonumber    &+\sum_{\beta>\gamma}^{N_\mathrm{M}} \sum_{k=1}^3 \int d \mathbf{r} \int_0^{\infty} d\omega \, C^{\mathrm{E}_{\gamma \beta},\left\{1_k\right\}}(\mathbf{r}, \omega, t') \\
   &\quad \times e^{-i W^{\mathrm{E}_{\gamma \beta},\{1\}}(\omega) t'} \ket{\mathrm{E}_\beta} \ket{\left\{1_k(\mathbf{r}, \omega)\right\} } \\
\nonumber    &+\sum_{\alpha<\gamma}^{N_\mathrm{M}} \sum_{k=1}^3 \int d \mathbf{r} \int_0^{\infty} d\omega \, C^{\mathrm{E}_{\alpha \gamma },\left\{1_k\right\}}(\mathbf{r}, \omega, t')\\
&\quad \times e^{-i W^{\mathrm{E}_{\alpha \gamma },\{1\}}(\omega) t'} \ket{\mathrm{E}_\alpha} \ket{\left\{1_k(\mathbf{r}, \omega)\right\} }.
\end{align}

On the other hand, the operator $ \hat{\sigma}_\gamma^{(+)}$ acting on the state $\ket{\Psi(t')}$ in Eq.~(\ref{Eq:Append_Sigma_ket_pm}) can be evaluated as
\begin{align}
\nonumber
    \hat{\sigma}_\gamma^{(+)} \ket{\Psi(t')} = & \sum_{\alpha<\gamma}^{N_\mathrm{M}} C^{\mathrm{E}_\alpha,\{0\}}(t') e^{-i W^{\mathrm{E}_\alpha,\{0\}} t'}\ket{\mathrm{E}_{\alpha\gamma}} \ket{\left\{0\right\}} \\
\nonumber
    &+ \sum_{\alpha>\gamma}^{N_\mathrm{M}} C^{\mathrm{E}_\alpha,\{0\}}(t') e^{-i W^{\mathrm{E}_\alpha,\{0\}} t'}\ket{\mathrm{E}_{\gamma\alpha}} \ket{\left\{0\right\}}  \\
\nonumber
    &+ \sum_{k=1}^3 \int {d} \mathbf{r} \int_0^{\infty} {d} \omega C^{\mathrm{G},\left\{1_k\right\}}(\mathbf{r}, \omega, t') \\
    &\quad \times e^{-i W^{\mathrm{G},\{1\}}(\omega) t'} \ket{\mathrm{E}_\gamma}\ket{\left\{1_k(\mathbf{r}, \omega)\right\}}. 
\label{Eq:Append_sigma(+)}
\end{align}

Note that we have disregarded three-molecule excitation states (e.g., $\ket{\mathrm{E}_{\gamma\alpha\beta}} = \hat{\sigma}^{(+)}_\gamma \hat{\sigma}^{(+)}_\alpha\hat{\sigma}^{(+)}_\beta \ket{\mathrm{G}}$) in Eq.~(\ref{Eq:Append_sigma(+)}) as they are not part of our state vector in Eq.~(\ref{Eq:State_Vector}). Eventually, Eq.~(\ref{Eq:Append_Sigma_ket_pm}) becomes:
\begin{widetext}
\begin{align}
\nonumber
 \left( \hat{\sigma}_\gamma^{(+)} + \hat{\sigma}_\gamma^{(-)} \right) \ket{\Psi(t')}  &=  \sum_{\alpha\neq\gamma}^{N_\mathrm{M}} C^{\mathrm{E}_\alpha,\{0\}}(t') e^{-i W^{\mathrm{E}_\alpha,\{0\}} t'}\ket{\mathrm{E}_{\alpha\gamma}} \ket{\left\{0\right\}} 
 + \sum_{k=1}^3 \int {d} \mathbf{r} \int_0^{\infty} {d} \omega C^{\mathrm{G},\left\{1_k\right\}}(\mathbf{r}, \omega, t')  e^{-i W^{\mathrm{G},\{1\}}(\omega) t'} \ket{\mathrm{E}_\gamma}\ket{\left\{1_k(\mathbf{r}, \omega)\right\}} \\
\nonumber
    &\quad +C^{\mathrm{E}_\gamma,\{0\}}(t') e^{-i W^{\mathrm{E}_\gamma,\{0\}} t'}\ket{\mathrm{G}} \ket{\left\{0\right\}} +\sum_{\beta>\gamma}^{N_\mathrm{M}} \sum_{k=1}^3 \int d \mathbf{r} \int_0^{\infty} d\omega \, C^{\mathrm{E}_{\gamma \beta},\left\{1_k\right\}}(\mathbf{r}, \omega, t')  e^{-i W^{\mathrm{E}_{\gamma \beta},\{1\}}(\omega) t'} \ket{\mathrm{E}_\beta} \ket{\left\{1_k(\mathbf{r}, \omega)\right\} } \\
    &\quad +\sum_{\alpha<\gamma}^{N_\mathrm{M}} \sum_{k=1}^3 \int d \mathbf{r} \int_0^{\infty} d\omega \, C^{\mathrm{E}_{\alpha \gamma },\left\{1_k\right\}}(\mathbf{r}, \omega, t')  e^{-i W^{\mathrm{E}_{\alpha \gamma },\{1\}}(\omega) t'} \ket{\mathrm{E}_\alpha} \ket{\left\{1_k(\mathbf{r}, \omega)\right\} } .
\label{Eq:App_Sigma_KetStates}
\end{align}
\end{widetext}

Next, we need to evaluate $e^{i\hat{H}_\mathrm{T}t'/\hbar}  \left( \hat{\sigma}_\gamma^{(+)} + \hat{\sigma}_\gamma^{(-)} \right) \ket{\Psi(t')}$ in Eq.~(\ref{Eq:Append_two_time_dipole}), which can be approximated as,
\begin{align}
\nonumber
    &e^{\frac{i \hat{H}_\mathrm{T} t'}{\hbar}  }  \left( \hat{\sigma}_\gamma^{(+)} + \hat{\sigma}_\gamma^{(-)} \right) \ket{\Psi(t')} \\
    &\approx \left\{ \mathrm{\hat{I}} +  \frac{i t'}{\hbar}  \hat{H}_\mathrm{T}\right\}  \left( \hat{\sigma}_\gamma^{(+)} + \hat{\sigma}_\gamma^{(-)} \right) \ket{\Psi(t')} .
\label{Eq:App_Further_Approx}
\end{align}

We find that $\hat{H}_\mathrm{T}$ acts on three types of states in Eq.~(\ref{Eq:App_Sigma_KetStates}), with the first type becoming:
\begin{align}
    &\hat{H}_\mathrm{T} \ket{\mathrm{G}} \ket{\left\{0\right\}} = \left(\hat{H}_\mathrm{M}+\hat{H}_\mathrm{P}\right)\ket{\mathrm{G}} \ket{\left\{0\right\}} + \hat{H}_\mathrm{M-P}\ket{\mathrm{G}} \ket{\left\{0\right\}} \\
    &\quad = 0  -\sum_{\alpha=1}^{N_\mathrm{M}} \hat{\pmb{\mu}}_\alpha \cdot \hat{\mathbf{E}}(\mathbf{r}_\alpha) \ket{\mathrm{G}} \ket{\left\{0\right\}} \\
    &\quad  =- \sum_{\alpha=1}^{N_\mathrm{M}} \sum_{j,k=1}^3 \int \mathrm{d}\mathbf{r}   \int_0^\infty d\omega \; {{\mu}}_{\alpha,j} \overline{\overline{\mathrm{g}}}_{jk}^*(\mathbf{r}_\alpha,\mathbf{r},\omega)  \ket{\mathrm{E}_\alpha} \ket{\left\{1_k(\mathbf{r},\omega)\right\}}.
\end{align}

For the second type of states, $\hat{H}_\mathrm{T}$ acts as follows:
\begin{align}
\nonumber
   &\hat{H}_\mathrm{T} \ket{\mathrm{E}_\alpha} \ket{\left\{1_k(\mathbf{r},\omega)\right\}} \\
\nonumber
   &= \hbar \left( \omega_\alpha + \omega \right)\ket{\mathrm{E}_\alpha} \ket{\left\{1_k(\mathbf{r},\omega)\right\}} \\
   &\quad -\sum_{\alpha'=1}^{N_\mathrm{M}} \hat{\pmb{\mu}}_{\alpha'} \cdot \hat{\mathbf{E}}(\mathbf{r}_{\alpha'}) \ket{\mathrm{E}_\alpha} \ket{\left\{1_k(\mathbf{r},\omega)\right\}} \\
\nonumber
 &=\hbar \left( \omega_\alpha + \omega \right)\ket{\mathrm{E}_\alpha} \ket{\left\{1_k(\mathbf{r},\omega)\right\}}- \sum_{j=1}^3 {{\mu}}_{\alpha,j} \overline{\overline{\mathrm{g}}}_{jk}(\mathbf{r}_{\alpha},\mathbf{r},\omega) \ket{\mathrm{G}} \ket{\left\{0\right\}}  \\
\nonumber
  &\quad  -\sum_{\alpha'>\alpha}^{N_\mathrm{M}}  \sum_{j=1}^3 {{\mu}}_{\alpha',j} \overline{\overline{\mathrm{g}}}_{jk}(\mathbf{r}_{\alpha'},\mathbf{r},\omega) \ket{\mathrm{E}_{\alpha\alpha'}} \ket{\left\{0\right\}} \\
  &\quad  -\sum_{\alpha'<\alpha}^{N_\mathrm{M}}  \sum_{j=1}^3 {{\mu}}_{\alpha',j} \overline{\overline{\mathrm{g}}}_{jk}(\mathbf{r}_{\alpha'},\mathbf{r},\omega) \ket{\mathrm{E}_{\alpha'\alpha}} \ket{\left\{0\right\}}.
\label{Eq:App_Eq_No_2Photon}
\end{align}

Note that we neglect two-polariton states (e.g., $\hat{f}_{k'}^\dagger (\mathbf{r}',\omega')\hat{f}_k^\dagger (\mathbf{r},\omega)\ket{\left\{0\right\}} $) in Eq.~(\ref{Eq:App_Eq_No_2Photon}) as they are not part of our state vector in Eq.~(\ref{Eq:State_Vector}). For the third type of states, $\hat{H}_\mathrm{T}$ acts as follows:
\begin{align}
\nonumber
    &\hat{H}_\mathrm{T} \ket{\mathrm{E}_{\alpha\gamma}} \ket{\left\{ 0 \right\}} \\ 
\nonumber
    &=  \hbar\left(\omega_\alpha + \omega_\gamma \right) \ket{\mathrm{E}_{\alpha\gamma}} \ket{\left\{ 0 \right\}} - \sum_{\beta=1}^{N_\mathrm{M}} \hat{\sigma}_\beta^{(-)}  \sum_{j,k=1}^3 \int \mathrm{d}\mathbf{r}   \int_0^\infty d\omega \; {{\mu}}_{\beta,j} \\
    &\quad  \times \overline{\overline{\mathrm{g}}}_{jk}^*(\mathbf{r}_\beta,\mathbf{r},\omega)  \hat{f}_{k}^\dagger(\mathbf{r},\omega) \ket{\mathrm{E}_{\alpha\gamma}} \ket{\left\{ 0 \right\}} \\
\nonumber
&=  \hbar\left(\omega_\alpha + \omega_\gamma \right) \ket{\mathrm{E}_{\alpha\gamma}} \ket{\left\{ 0 \right\}} \\
\nonumber
&\quad - \sum_{j,k=1}^3 \int \mathrm{d}\mathbf{r}   \int_0^\infty d\omega \; {{\mu}}_{\gamma,j} \overline{\overline{\mathrm{g}}}_{jk}^*(\mathbf{r}_\gamma,\mathbf{r},\omega) \ket{\mathrm{E}_{\alpha}}\ket{\left\{ 1_k(\mathbf{r},\omega) \right\}} \\
& \quad - \sum_{j,k=1}^3 \int \mathrm{d}\mathbf{r}   \int_0^\infty d\omega \; {{\mu}}_{\alpha,j} \overline{\overline{\mathrm{g}}}_{jk}^*(\mathbf{r}_\alpha,\mathbf{r},\omega)  \ket{\mathrm{E}_{\gamma}}   \ket{\left\{ 1_k(\mathbf{r},\omega) \right\}} .
\label{Eq:App_No_3_Exciton}
\end{align}
 
Note that we have disregarded three-molecule excitation states in Eq.~(\ref{Eq:App_No_3_Exciton}). We also disregarded the states $\ket{\mathrm{E}_\alpha} 
\ket{\left\{1_k(\mathbf{r},\omega)\right\} } $, $\ket{\mathrm{E}_\beta} 
\ket{\left\{1_k(\mathbf{r},\omega)\right\} } $, and $\ket{\mathrm{E}_\gamma} 
\ket{\left\{1_k(\mathbf{r},\omega)\right\} } $ in Eq.~(\ref{Eq:App_Sigma_KetStates}) as they are not part of our state vector. Consequently, we approximate Eq.~(\ref{Eq:App_Further_Approx}) as
\begin{align}
\nonumber
&e^{i\hat{H}_\mathrm{T}t'/\hbar}  \left( \hat{\sigma}_\gamma^{(+)} + \hat{\sigma}_\gamma^{(-)} \right) \ket{\Psi(t')} \\
\nonumber
&\approx  
\sum_{\alpha\neq\gamma}^{N_\mathrm{M}} C^{\mathrm{E}_\alpha,\{0\}}(t') e^{+i W^{\mathrm{E}_\gamma,\{0\}} t'}\ket{\mathrm{E}_{\alpha\gamma}} \ket{\left\{0\right\}} \\
& \quad + C^{\mathrm{E}_\gamma,\{0\}}(t') e^{-i W^{\mathrm{E}_\gamma,\{0\}} t'}\ket{\mathrm{G}} \ket{\left\{0\right\}}  .
\label{Eq:Append_exp_sigma_Phi}
\end{align}

After substituting Eqs.~(\ref{Eq:App_Sigma_KetStates}) and (\ref{Eq:Append_exp_sigma_Phi}) into Eq.~(\ref{Eq:Two_time_Schrodinger}), and then Eq.~(\ref{Eq:Two_time_Schrodinger}) into Eq.~(\ref{Eq:Lineshape_Func}), the expression for the LF becomes
\begin{align}
\nonumber  &\mathcal{L}^{(N_\mathrm{M})}_{\alpha,\beta}(\omega_\mathrm{D}) 
   \\
\nonumber
&= \int_0^{\infty} dt'' \int_0^{\infty} dt' \; e^{-i\omega_\mathrm{D} (t''-t')}  \left< \Psi(t'') \right|   \left(\hat{\sigma}_{\beta}^{(-)}+ \hat{\sigma}_{\beta}^{(+)}\right)  \\ 
   & \times e^{i\hat{H}_\mathrm{T}(t'-t'')/\hbar}
    \left(\hat{\sigma}_{\alpha}^{(-)}+ \hat{\sigma}_{\alpha}^{(+)}\right) \left| \Psi(t') \right> \\
\nonumber
&\approx   \int_0^{\infty} dt'' \int_0^{\infty} dt' \; e^{-i\omega_\mathrm{D} (t''-t')} C^{\mathrm{E}_\beta,\{0\}*}(t'') \\
\nonumber
 &\times  e^{i W^{\mathrm{E}_\beta,\{0\}} t''} C^{\mathrm{E}_\alpha,\{0\}}(t') e^{-i W^{\mathrm{E}_\alpha,\{0\}} t'} \\
\nonumber
 &+ \int_0^{\infty} dt'' \int_0^{\infty} dt' \; e^{-i\omega_\mathrm{D} (t''-t')}  \sum_{\beta'\neq\beta}^{N_\mathrm{M}} C^{\mathrm{E}_{\beta'},\{0\}*}(t'') \\
 &\times e^{-i W^{\mathrm{E}_\beta,\{0\}} t''}\sum_{\alpha' \neq\alpha}^{N_\mathrm{M}} C^{\mathrm{E}_{\alpha'},\{0\}}(t') e^{+i W^{\mathrm{E}_\alpha,\{0\}} t'}\braket{\mathrm{E}_{\beta'\beta}}{\mathrm{E}_{\alpha'\alpha}} ,  
\label{Eq:Append_LF_temp1}
\end{align}

Because of $\braket{\mathrm{E}_{\beta'\beta}}{\mathrm{E}_{\alpha'\alpha}} = \bra{\mathrm{G}} \hat{\sigma}_{\beta}^{(-)} \hat{\sigma}_{\beta'}^{(-)} \hat{\sigma}_{\alpha'}^{(+)} \hat{\sigma}_{\alpha}^{(+)} \ket{\mathrm{G}} 
=       \delta_{\beta,\alpha'} \delta_{\alpha,\beta'} + \delta_{\beta,\alpha} \delta_{\alpha',\beta'}  + 2\delta_{\beta,\beta'} \delta_{\alpha',\beta'}\delta_{\alpha,\beta'}  $, Eq.~(\ref{Eq:Append_LF_temp1}) reduces to 
\begin{widetext}

\begin{align}
\nonumber
    \mathcal{L}^{(N_\mathrm{M})}_{\alpha,\beta}(\omega_\mathrm{D}) =&  \int_0^{\infty} dt'' \int_0^{\infty} dt' \; e^{-i\omega_\mathrm{D} (t''-t')} C^{\mathrm{E}_\beta,\{0\}*}(t'') e^{i W^{\mathrm{E}_\beta,\{0\}} t''} C^{\mathrm{E}_\alpha,\{0\}}(t') e^{-i W^{\mathrm{E}_\alpha,\{0\}} t'} \\
\nonumber
    &+ \int_0^{\infty} dt'' \int_0^{\infty} dt' \; e^{-i\omega_\mathrm{D} (t''-t')}   C^{\mathrm{E}_{\alpha},\{0\}*}(t'') e^{-i W^{\mathrm{E}_\beta,\{0\}} t''}  C^{\mathrm{E}_{\beta},\{0\}}(t') e^{+i W^{\mathrm{E}_\alpha,\{0\}} t'}   \\
    &+ \int_0^{\infty} dt'' \int_0^{\infty} dt' \; e^{-i\omega_\mathrm{D} (t''-t')}  \sum_{\beta'\neq\beta}^{N_\mathrm{M}} C^{\mathrm{E}_{\beta'},\{0\}*}(t'') e^{-i W^{\mathrm{E}_\beta,\{0\}} t''}  C^{\mathrm{E}_{\beta'},\{0\}}(t') e^{+i W^{\mathrm{E}_\alpha,\{0\}} t'} \delta_{\beta,\alpha} \\
\nonumber
    =&  \left[ \mathfrak{L}_{t''} \left\{C^{\mathrm{E}_{\beta},\{0\}}(t'') \right\}(i(W^{\mathrm{E}_{\beta},\{0\}}-\omega_\mathrm{D}) ) \right]^*   \mathfrak{L}_{t'} \left\{C^{\mathrm{E}_{\alpha},\{0\}}(t') \right\}(i(W^{\mathrm{E}_{\alpha},\{0\}}-\omega_\mathrm{D}) ) \\
\nonumber
    &+  \left[ \mathfrak{L}_{t''} \left\{C^{\mathrm{E}_{\alpha},\{0\}}(t'') \right\}(-i(W^{\mathrm{E}_{\beta},\{0\}}+\omega_\mathrm{D}) ) \right]^*   \mathfrak{L}_{t'} \left\{C^{\mathrm{E}_{\beta},\{0\}}(t') \right\}(-i(W^{\mathrm{E}_{\alpha},\{0\}}+\omega_\mathrm{D}) )\\
    &+ \sum_{\beta'\neq \beta}^{N_\mathrm{M}} \left[ \mathfrak{L}_{t''} \left\{C^{\mathrm{E}_{\beta'},\{0\}}(t'') \right\}(-i(W^{\mathrm{E}_{\beta},\{0\}}+\omega_\mathrm{D}) ) \right]^*   \mathfrak{L}_{t'} \left\{C^{\mathrm{E}_{\beta'},\{0\}}(t') \right\}(-i(W^{\mathrm{E}_{\alpha},\{0\}}+\omega_\mathrm{D}) ) \delta_{\alpha,\beta} .
\label{Eq_App_Correlation_Dipole}
\end{align}

The last two terms in Eq.~(\ref{Eq_App_Correlation_Dipole}) contribute most when $\omega_\mathrm{D} \approx -W^{\mathrm{E}_{\beta},\{0\}} <0 $ and $\omega_\mathrm{D} \approx -W^{\mathrm{E}_{\alpha},\{0\}} <0 $; however, since the detection frequency $\omega_\mathrm{D}$ must be larger than zero, we drop the last two terms. Consequently, we approximate the LF in Eq.~(\ref{Eq:Lineshape_Func}) as:
\begin{align}
    \mathcal{L}^{(N_\mathrm{M})}_{\alpha,\beta}(\omega_\mathrm{D}) \approx  \left[ \mathfrak{L}_{t''} \left\{C^{\mathrm{E}_{\beta},\{0\}}(t'') \right\}(i(W^{\mathrm{E}_{\beta},\{0\}}-\omega_\mathrm{D}) ) \right]^*   \mathfrak{L}_{t'} \left\{C^{\mathrm{E}_{\alpha},\{0\}}(t') \right\}(i(W^{\mathrm{E}_{\alpha},\{0\}}-\omega_\mathrm{D}) ) .
\end{align}

\section{Reflection Coefficients and Auxiliary Tensors in Far-Field Region}
\label{App_Reflection_FarField}

Based on previous studies\cite{chew1995waves,novotny2012principles}, the $\sigma$-polarized reflection coefficient for the interface between the zeroth medium and the first medium (recall Eq.~(\ref{dielectrics})) can be expressed as:
\begin{align}
    r_{\sigma,01}(\bar q,\omega_\mathrm{D})=
    \begin{cases}
        \displaystyle
        \frac{K_{\mathrm{z},0}-K_{\mathrm{z},1}}{K_{\mathrm{z},0}+K_{\mathrm{z},1}},
        \qquad \sigma=\mathrm{s},\\
        \\
        \displaystyle
        \frac{\epsilon_{\mathrm{r},1}K_{\mathrm{z},0}-\epsilon_{\mathrm{r},0}K_{\mathrm{z},1}}{\epsilon_{\mathrm{r},1}K_{\mathrm{z},0}+\epsilon_{\mathrm{r},0}K_{\mathrm{z},1}},
        \qquad \sigma=\mathrm{p},
    \end{cases}
\label{App_Eq_Reflection_Coeff}
\end{align}
where $K_{\mathrm{z},j} = \sqrt{ \epsilon_{\mathrm{r},j}(\omega_\mathrm{D})k_0^2 - \bar q^2 } $ is the z-component wavevector in the $j$-th medium, and $\bar{q} = k_0 \sin{\bar{\theta}}$ is the magnitude of the in-plane wavevector. The far-field auxiliary tensors $\overline{\overline{\mathbf{M}}}_{\sigma,\mathrm{far}}(\bar\theta,\bar\phi)$ can be expressed in the Cartesian coordinate as\cite{lee_2021}:
\begin{align}
    &\overline{\overline{\mathbf{M}}}_{\mathrm{s,far}}(\bar\theta,\bar\phi) =   \begin{bmatrix}
        1-\cos^2{\bar{\phi}} & -\frac{1}{2}\sin{2\bar{\phi}} & 0\\
        -\frac{1}{2}\sin{2\bar{\phi}} & 1-\sin^2{\bar{\phi}} & 0\\
        0 & 0 & 0 \\
    \end{bmatrix} ,  \\ &\overline{\overline{\mathbf{M}}}_{\mathrm{p,far}}(\bar\theta,\bar\phi)  = \begin{bmatrix}
    -\cos^2{\bar\theta}\cos^2{\bar\phi} & -\frac{1}{2}\cos^2{\bar\theta}\sin{2\bar\phi} & -\frac{1}{2}\sin{2\bar\theta}\cos{\bar\phi}\\
    -\frac{1}{2}\cos^2{\bar\theta}\sin{2\bar\phi} & -\cos^2{\bar\theta}\sin^2{\bar\phi} & -\frac{1}{2}\sin{2\bar\theta}\sin{\bar\phi}\\
    \frac{1}{2}\sin{2\bar\theta}\cos{\bar\phi} & \frac{1}{2}\sin{2\bar\theta}\sin{\bar\phi} & \sin^2\bar\theta
    \end{bmatrix},
\label{Eq:M_p_far}
\end{align}
where $\bar{\phi} = \arctan{(y/x)}$ and $\bar{\theta} = \arcsin{(\rho/\bar{R})}$ are the azimuthal and polar angle in the spherical coordinate, respectively. $\rho=\bar{R}\sin{\bar\theta}=\sqrt{x^2+y^2}$ is the axial radius. Here, we would like to emphasize that the azimuthal and the polar angle of the detector are $\phi_\mathrm{D}=\arctan(y/x)$ and $\theta_\mathrm{D}=\arcsin(\rho/R)$, respectively. Therefore, $\phi_\mathrm{D}=\bar{\phi}$ and $\theta_\mathrm{D}\approx \bar{\theta}$ when $R\approx \bar{R}$, which is a valid approximation for our setup.

\section{Validity of $\overline{\overline{\mathbf{G}}}(\mathbf{R}_\mathrm{D},\mathbf{r}_{2(3)},\omega_\mathrm{D}) \approx \overline{\overline{\mathbf{G}}}(\mathbf{R}_\mathrm{D},\mathbf{r}_{1},\omega_\mathrm{D})$ for Closely Aligned Molecules}
\label{App_EEF_r2}

The dyadic Green's function in Eq.~(\ref{Eq:GEEF_Reduced}) for the second molecule can be expressed as
\begin{align}  \overline{\overline{\mathbf{G}}}(\mathbf{R}_\mathrm{D},\mathbf{r}_2,\omega_\mathrm{D}) \approx & \, \overline{\overline{\mathbf{G}}}_{0,\mathrm{far}}(\mathbf{R}_\mathrm{D},\mathbf{r}_2,\omega_\mathrm{D}) +   \overline{\overline{\mathbf{G}}}_{\mathrm{s,far}}(\mathbf{R}_\mathrm{D},\mathbf{r}_2,\omega_\mathrm{D}) +\overline{\overline{\mathbf{G}}}_{\mathrm{p,far}}(\mathbf{R}_\mathrm{D},\mathbf{r}_2,\omega_\mathrm{D}),
\end{align}
where the position of the second molecule and the detector are $\mathbf{r}_\mathrm{2} =(d,0,h)$ and $\mathbf{R}_\mathrm{D} = (x,y,z)$, respectively. Since the free-space dyadic Green's function for the first molecule at $\mathbf{r}_\mathrm{1} = (0,0,h)$ can be expressed as (recall Eq.~(\ref{Eq:Green_Func_Vac})),
\begin{align}
    \tensorg_{0,\mathrm{far}}(\mathbf{R}_\mathrm{D},\mathbf{r}_1,\omega_\mathrm{D}) &= \frac{e^{ik_0 R}}{4\pi R } \left( \begin{bmatrix}
        1 & 0 & 0\\
        0 & 1 & 0\\
        0 & 0 & 1
    \end{bmatrix} - \frac{1}{R } \begin{bmatrix}
        x^2 & xy & x(z-h)\\
        xy & y^2 & y(z-h)\\
        x(z-h) & y(z-h) & (z-h)^2
    \end{bmatrix} \right)  = \tensorg_{0,\mathrm{far}}(x,y,z,\omega_\mathrm{D}),
    \label{Eq:g0_far_D1}
\end{align}
we can perform a Taylor expansion of the free-space dyadic Green's function for the second molecule $ \tensorg_{0,\mathrm{far}}(\mathbf{R}_\mathrm{D},\mathbf{r}_2,\omega_\mathrm{D}) = \tensorg_{0,\mathrm{far}}(x-d,y,z,\omega_\mathrm{D}) \approx \tensorg_{0,\mathrm{far}}(x,y,z,\omega_\mathrm{D}) - \left( \partial \tensorg_{0,\mathrm{far}}(x,y,z,\omega_\mathrm{D}) / \partial x\right) d  $, and obtain: 

\begin{align}
\nonumber
    \tensorg_{0,\mathrm{far}}(x-d,y,z,\omega_\mathrm{D}) 
    &\approx \tensorg_{0,\mathrm{far}}(\mathbf{R}_\mathrm{D},\mathbf{r}_1,\omega_\mathrm{D}) -  \frac{e^{ik_0 R}}{4\pi R } \left(\overline{\overline{\mathbf{I}}}_3-\hat{\mathbf{e}}_{ {R} }\otimes\hat{\mathbf{e}}_{ {R} } \right) \left(\frac{-x}{R^2} + \frac{i k_0 x}{R }  \right) d \\
    &\quad +  \frac{e^{ik_0 R }}{4\pi R^4 }  \begin{bmatrix}
        2x(y^2+(z-h)^2) & y(-x^2+y^2+(z-h)^2) & (z-h)(-x^2+y^2+(z-h)^2)\\
        y(-x^2+y^2+(z-h)^2) & -2xy^2 & -2xy(z-h)\\
        (z-h)(-x^2+y^2+(z-h)^2) & -2xy(z-h) & -2x(z-h)^2
    \end{bmatrix}  \frac{d}{R}.
\label{Eq:App_G0_diff}
\end{align}

In spherical coordinates, the position of the detector can be expressed as $\mathbf{R}_\mathrm{D} = R (\sin{\theta_\mathrm{D}}\cos{\phi_\mathrm{D}},\sin{\theta_\mathrm{D}}\sin{\phi_\mathrm{D}},\cos{\theta_\mathrm{D}}) + (0,0,h)$, where $\theta_\mathrm{D}$ and $\phi_\mathrm{D}$ are the polar and azimuthal angles of the detector, respectively. Therefore, $x =R\sin{\theta_\mathrm{D}}\cos{\phi_\mathrm{D}}$, $y=R \sin{\theta_\mathrm{D}}\sin{\phi_\mathrm{D}}$, $z = h + R \cos{\theta_\mathrm{D}} $, and Eq.~(\ref{Eq:App_G0_diff}) becomes
\small{
\begin{align}
\nonumber
&\tensorg_{0,\mathrm{far}}(\mathbf{R}_\mathrm{D},\mathbf{r}_2,\omega_\mathrm{D}) \approx  \;   \tensorg_{0,\mathrm{far}}(\mathbf{R}_\mathrm{D},\mathbf{r}_1,\omega_\mathrm{D}) \left[ 1 -    \left(\frac{-d}{R } + i k_0 d \right) \sin{\theta_\mathrm{D}}\cos{\phi_\mathrm{D}} \right] \\
    & +  \frac{e^{ik_0 R }}{4\pi R }
     \begin{bmatrix}
        2\sin{\theta_\mathrm{D}}\cos{\phi_\mathrm{D}}(\sin^2{\theta_\mathrm{D}}\sin^2{\phi_\mathrm{D}}+\cos^2{\theta_\mathrm{D}}) & \sin{\theta_\mathrm{D}}\sin{\phi_\mathrm{D}}(\cos^2{\theta_\mathrm{D}}-\sin^2{\theta_\mathrm{D}}\cos{2\phi_\mathrm{D}}) & \cos^3{\theta_\mathrm{D}}-\sin^2{\theta_\mathrm{D}}\cos{\theta_\mathrm{D}}\cos{2\phi_\mathrm{D}})\\
       \sin{\theta_\mathrm{D}}\sin{\phi_\mathrm{D}}(\cos^2{\theta_\mathrm{D}}-\sin^2{\theta_\mathrm{D}}\cos{2\phi_\mathrm{D}}) & -2\sin^3{\theta_\mathrm{D}}\sin^2{\phi_\mathrm{D}} \cos{\phi_\mathrm{D}} & -2\sin^2{\theta_\mathrm{D}}\cos{\phi_\mathrm{D}} \sin{\phi_\mathrm{D}}\cos{\theta_\mathrm{D}}\\
       \cos^3{\theta_\mathrm{D}}-\sin^2{\theta_\mathrm{D}}\cos{\theta_\mathrm{D}}\cos{2\phi_\mathrm{D}}) & -2\sin^2{\theta_\mathrm{D}}\cos{\phi_\mathrm{D}} \sin{\phi_\mathrm{D}}\cos{\theta_\mathrm{D}} & -2\sin{\theta_\mathrm{D}}\cos{\phi_\mathrm{D}}\cos^2{\theta_\mathrm{D}}
    \end{bmatrix} \frac{d}{R } .
\end{align}}
\normalsize

Clearly, when the intermolecular distance $d$ is much smaller than both the distance between the molecule and the detector ($d/R  \ll 1$) and the wavelength of the photon ($k_0 d = 2\pi d/\lambda_\mathrm{D} \ll 1 $), the two free-space dyadic Green's function are approximately equal, i.e., $\tensorg_{0,\mathrm{far}}(\mathbf{R}_\mathrm{D},\mathbf{r}_2,\omega_\mathrm{D}) \approx   \tensorg_{0,\mathrm{far}}(\mathbf{R}_\mathrm{D},\mathbf{r}_1,\omega_\mathrm{D})$.

Similarly, we can perform Taylor expansions of the $\sigma$-polarized dyadic Green's function for the second molecule:
\begin{align}
\label{Eq:Apend_GSigma}    \tensorg_{\sigma,\mathrm{far}}(x-d,y,z,\omega_\mathrm{D}) \approx \tensorg_{\sigma,\mathrm{far}}(x,y,z,\omega_\mathrm{D}) - \frac{ \partial \tensorg_{\sigma,\mathrm{far}}(x,y,z,\omega_\mathrm{D}) }{\partial x} d ,
\end{align}
where $\tensorg_{\sigma,\mathrm{far}}(x,y,z,\omega_\mathrm{D}) = \tensorg_{\sigma,\mathrm{far}}(\mathbf{R}_\mathrm{D},\mathbf{r}_1,\omega_\mathrm{D}) $ is expressed in a spherical coordinate, as shown in Eq.~(\ref{Eq:Gsigmafar}). Therefore, we need to apply the chain rule to evaluate its derivative:
\begin{align}
\nonumber
     \frac{ \tensorg_{\sigma,\mathrm{far}}(x,y,z,\omega_\mathrm{D}) }{\partial x} = &\frac{\partial}{\partial x} \left(  \frac{ e^{ik_0\bar{R}}}{4\pi\bar{R}} \right)  r_{\sigma,01}(\bar{q},\omega)
    \overline{\overline{\mathbf{M}}}_{\sigma\mathrm{,far}}(\bar\theta,\bar\phi) + \frac{\partial}{\partial x} \left(   r_{\sigma,01}(\bar{q},\omega) \right) \frac{ e^{ik_0\bar{R}}}{4\pi\bar{R}} 
    \overline{\overline{\mathbf{M}}}_{\sigma\mathrm{,far}}(\bar\theta,\bar\phi) \\
    & + \frac{\partial}{\partial x} \left(  \overline{\overline{\mathbf{M}}}_{\sigma\mathrm{,far}}(\bar\theta,\bar\phi)  \right) \frac{ e^{ik_0\bar{R}}}{4\pi\bar{R}} r_{\sigma,01}(\bar{q},\omega),
\label{Eq:Append_Deri_Gsigma}
\end{align}
with
\begin{center}
\begin{align}
&\frac{\partial}{\partial x} \left(  \frac{ e^{ik_0\bar{R}}}{4\pi\bar{R}} \right) =  \frac{ e^{ik_0\bar{R}}}{4\pi\bar{R}} \sin{\bar \theta}\cos{\bar \phi}  \left(\frac{-1}{\bar R } + i k_0  \right), \\
   &\frac{\partial}{\partial x} \left(   r_{\mathrm{s},01}(\bar{q},\omega) \right) = \frac{ -2\bar{q}k_0\left(K^{-1}_{\mathrm{z},0}-K^{-1}_{\mathrm{z},1}\right)}{K_{\mathrm{z},0}+K_{\mathrm{z},1}}   \frac{ \cos{\bar\phi} \cos^2{\bar\theta}}{\bar{R}} ,\\
  &\frac{\partial}{\partial x} \left(   r_{\mathrm{p},01}(\bar{q},\omega) \right) = \frac{ 2\bar{q} k_0^3 \epsilon_{\mathrm{r},0}\epsilon_{\mathrm{r},1}(\epsilon_{\mathrm{r},0}-\epsilon_{\mathrm{r},1}) }{  K_{\mathrm{z},0}K_{\mathrm{z},1}\left(\epsilon_{\mathrm{r},1}K_{\mathrm{z},0}+\epsilon_{\mathrm{r},0}K_{\mathrm{z},1}\right)^2 }  \frac{  \cos{\bar\phi} \cos^2{\bar\theta}}{\bar{R}},\\
&\frac{\partial}{\partial x} \left( \overline{\overline{\mathbf{M}}}_{\mathrm{s,far}}(\bar\theta,\bar\phi) \right)  =  \frac{ 1 }{ \rho } \begin{bmatrix}
        -2\cos{\bar{\phi}} \sin^2{\bar{\phi}}  &  \cos{2\bar{\phi}}\sin{\bar\phi}  & 0\\
           \cos{2\bar{\phi}}\sin{\bar\phi}  & 2\cos{\bar{\phi}} \sin^2{\bar{\phi}}   & 0\\
        0 & 0 & 0 \\
    \end{bmatrix}, 
\end{align}
\end{center}
\begin{align}
\nonumber
&\frac{\partial}{\partial x} \left( \overline{\overline{\mathbf{M}}}_{\mathrm{p,far}}(\bar\theta,\bar\phi) \right) \\
    &= \frac{1}{\rho} \begin{bmatrix}
    { 2\cos^2{\bar\theta}\cos^3{\bar\phi} \left( \sin{\bar\theta}-\tan^2{\bar\phi} \right) }   &  { \cos^2{\bar\theta} \left( \sin{\bar\theta}\sin{2\bar\phi}\cos{\bar\phi} + \sin{\bar\phi}\cos{2\bar\phi} \right)  }   &  {-\cos{\bar\theta}\left( \sin{\bar\theta}\sin^2{\bar\phi} + \cos{2\bar\theta}\cos^2{\bar\phi} \right)  }  \\
      { \cos^2{\bar\theta} \left( \sin{\bar\theta}\sin{2\bar\phi}\cos{\bar\phi} + \sin{\bar\phi}\cos{2\bar\phi} \right)  }  &   {2\cos^2{\bar\theta}\sin^2{\bar\phi}\cos{\bar\phi} \left( \sin{\bar\theta} + 1 \right)   }   &   {\cos{\bar\theta}\sin{\bar\phi}\cos{\bar\phi} \left( \sin{\bar\theta} - \cos{2\bar\theta} \right) }   \\
    {-\cos{\bar\theta}\left( \sin{\bar\theta}\sin^2{\bar\phi} + \cos{2\bar\theta}\cos^2{\bar\phi} \right)  }  &  {\cos{\bar\theta}\sin{\bar\phi}\cos{\bar\phi} \left( \sin{\bar\theta} - \cos{2\bar\theta} \right) }     &   { 2\cos^2{\bar\theta}\sin{\bar\theta}\cos{\bar\phi} } 
    \end{bmatrix}. 
\end{align}
\end{widetext}

Substituting Eq.~(\ref{Eq:Append_Deri_Gsigma}) into Eq.~(\ref{Eq:Apend_GSigma}), we find that when $d/\bar R  \ll 1$, $d/\rho \ll 1$, and $k_0 d \ll 1 $, the two $\sigma$-polarized dyadic Green's function are approximately equal, i.e., $\tensorg_{\sigma,\mathrm{far}}(\mathbf{R}_\mathrm{D},\mathbf{r}_2,\omega_\mathrm{D}) \approx   \tensorg_{\sigma,\mathrm{far}}(\mathbf{R}_\mathrm{D},\mathbf{r}_1,\omega_\mathrm{D})$. Given that $\rho \leq \bar R$, if $d/\rho \ll 1$, then $d/\bar R \ll 1$ is automatically satisfied. Thus, when the intermolecular distance $d$ is much smaller than the axial radius between the molecule and the detector (i.e., $d/\rho \ll 1$), and the wavelength of the photon (i.e., $k_0 d = 2\pi d/\lambda_\mathrm{D} \ll 1 $), the dyadic Green's function of the second molecule is approximately equal to that of the first molecule, i.e.,
\begin{align}
    \tensorg_{\mathrm{far}}(\mathbf{R}_\mathrm{D},\mathbf{r}_2,\omega_\mathrm{D}) \approx   \tensorg_{\mathrm{far}}(\mathbf{R}_\mathrm{D},\mathbf{r}_1,\omega_\mathrm{D}).
\end{align}

Similarly, for the third molecule, when $2d/\rho \ll 1$ and $2d k_0 \ll 1 $, the corresponding dyadic Green's function is approximately equal to that of the first molecule, i.e.,
\begin{align}
    \tensorg_{\mathrm{far}}(\mathbf{R}_\mathrm{D},\mathbf{r}_3,\omega_\mathrm{D}) \approx   \tensorg_{\mathrm{far}}(\mathbf{R}_\mathrm{D},\mathbf{r}_1,\omega_\mathrm{D}).
\end{align}

\section{Derivation of Analytical Solutions of LFs }
\label{Appendix_Analytical_Solution}

In this study, with all molecules identical and initially excited in a delocalized state, the analytical solutions of Eq.~(\ref{Eq:non_Hermitian_Schr_Eq}) for the monomer, dimer, and trimer systems are derived as follows:
\begin{align}
\label{Eq:APP_Analytical_Solution_2_Momer}
&\mathrm{Monomer:} \qquad     C^{\mathrm{E}_1,\{0\}}(t) = \exp\left(-\frac{\Gamma}{2}t \right), \\
&\mathrm{Dimer:} \qquad \begin{cases}
C^{\mathrm{E}_1,\{0\}}(t) =  \frac{1}{\sqrt{2}} \exp\left(-\frac{\hbar\Gamma + 2iV}{2\hbar} t\right) ,\\
C^{\mathrm{E}_2,\{0\}}(t) =C^{\mathrm{E}_1,\{0\}}(t) , 
\end{cases}
\label{Eq:APP_Analytical_Solution_2}\\
&\mathrm{Trimer:} \quad \begin{cases}
    C^{\mathrm{E}_1,\{0\}}(t) =  D_1 \exp\left(-\frac{\hbar\Gamma + i(V'+V_\mathrm{eff})}{2\hbar} t\right) \\
    \qquad \qquad \quad  - D_2 \exp\left(-\frac{\hbar\Gamma + i(V'-V_\mathrm{eff})}{2\hbar} t\right) ,   \\
    C^{\mathrm{E}_2,\{0\}}(t) = - D_1 D_3 \exp\left(-\frac{\hbar\Gamma + i(V'+V_\mathrm{eff})}{2\hbar} t\right) \\
    \qquad \qquad \quad  + D_2 D_4 \exp\left(-\frac{\hbar\Gamma + i(V'-V_\mathrm{eff})}{2\hbar} t\right) ,
  \\
C^{\mathrm{E}_3,\{0\}}(t) = C^{\mathrm{E}_1,\{0\}}(t) ,  
\end{cases}
\label{Eq:APP_Analytical_Solution_2_Trimer}
\end{align}
where the decay rates $\Gamma=\Gamma_\alpha$ are identical for all molecules in the three systems. The symbols $V$, $V'$, and $V_\mathrm{eff}$ are defined as: $V=V_{\mathrm{RDDI},12}$, $V'= V_{\mathrm{RDDI},13}$, and $V_\mathrm{eff} = \sqrt{8V^2+V'^2}$. Based on Eqs.~(\ref{Eq:Decay_Rate_Gamma}) and (\ref{Eq:RDDI}), the values of $V$, $V'$ and $V_\mathrm{eff}$ are calculated and summarized in Table~\ref{tab:parameters}. The symbols $D_1$, $D_2$, $D_3$, and $D_4$ are further defined as: $D_1 = {\sqrt{3}\left(2V+V'+V_\mathrm{eff}\right)}/{(6V_\mathrm{eff})}$, $D_2 = {\sqrt{3}\left(2V+V'-V_\mathrm{eff}\right)}/{(6V_\mathrm{eff})}$, $D_3 = {(V'-V_\mathrm{eff})}/{(2V)}$, and $D_4= {(V'+V_\mathrm{eff})}/{(2V)}$. The values of these parameters are summarized in Table~\ref{tab:parameters_App}:
\begin{table}[h]
    \centering
    \begin{tabular}{c | c | c | c }
        \toprule
        Parameter & Value & Parameter & Value \\
         \hline
    $D_1$ & $0.5055 + 0.0006i$ & $D_2$ & $-0.0719 + 0.0006i$ \\
        $D_3$ & $-1.3524+0.0031i$  &  $D_4$ & $1.4788+0.0034i$\\
        \toprule
    \end{tabular}
    \caption{Values of parameters $D_1$, $D_2$, $D_3$, and $D_4$ from Eq.~(\ref{Eq:APP_Analytical_Solution_2_Trimer}), given in units of meV. }
    \label{tab:parameters_App}
\end{table}

Substituting Eqs.~(\ref{Eq:APP_Analytical_Solution_2_Momer}) and (\ref{Eq:APP_Analytical_Solution_2}) into Eq.~(\ref{Eq:LF_11}), We derived the analytical expressions for the specific LFs of the dimer and trimer systems, as presented in Eqs.~(\ref{Eq:Analytical_Solution_LF3_1}) and (\ref{Eq:Analytical_Solution_LF3_2}) in the main text,
\begin{align}
\nonumber
  &\mathrm{Monomer}: \\
\label{Eq:Analytical_Solution_LF3_1_APP}
  &  \mathcal{L}^{(1)}(\omega_\mathrm{D}) = \frac{1}{ \left(\omega_\mathrm{D}-\omega_\mathrm{M} \right)^2+ \left(\frac{\Gamma}{2}\right)^2}, \\
\nonumber
  &\mathrm{Dimer}: \\
\label{Eq:Analytical_Solution_LF3_2_APP}
  &  \mathcal{L}^{(2)}(\omega_\mathrm{D}) = \frac{2}{ \left(\omega_\mathrm{D}-\omega_\mathrm{M}-\mathrm{Re}\{\frac{V}{\hbar}\} \right)^2+\left(\frac{\Gamma}{2}-\mathrm{Im}\{\frac{V}{\hbar}\}\right)^2}.
\end{align}
Substituting Eq.~(\ref{Eq:APP_Analytical_Solution_2_Trimer})
into Eq.~(\ref{Eq:LF_11}), we obtain the analytical expression for the LF of the trimer system as follows:
\begin{align}
\nonumber
&\mathcal{L}^{(3)}(\omega_\mathrm{D})\\
\nonumber
&= \left| \frac{D_1(2-D_3)}{i(\omega_1+\mathrm{Re}\{\frac{V'+V_\mathrm{eff}}{2\hbar}\} -\omega_\mathrm{D} ) + (\frac{\Gamma}{2} - \mathrm{Im}\{\frac{V'+V_\mathrm{eff}}{2\hbar} \} ) }
\right. \\
& \quad \left. +\frac{D_2(D_4-2)}{i(\omega_1+\mathrm{Re}\{ \frac{V'-V_\mathrm{eff}}{2\hbar}\} -\omega_\mathrm{D} ) + (\frac{\Gamma}{2} - \mathrm{Im}\{\frac{V'+V_\mathrm{eff}}{2\hbar} \} ) }
\right|^2 .
\label{APP_Eq_Lorentz_Trimer}
\end{align}
Since the second term corresponds to the blue-shifted peak at $\omega_\mathrm{D}\approx 3.665$ eV in Fig~\ref{Fig_4}, with negligible peak height, we omit it in Eq.~(\ref{APP_Eq_Lorentz_Trimer}) and obtain the simplified analytical expression for the LFs of the trimer system presented in the main text:
\begin{align}
\nonumber
    &\mathcal{L}^{(3)}(\omega_\mathrm{D}) \\ 
\nonumber
    &\approx \frac{ \abs{D_1}^2\abs{2-D_3}^2}{(\omega_1+\mathrm{Re}\{\frac{V'+V_\mathrm{eff}}{2\hbar}\} -\omega_\mathrm{D} )^2 + (\Gamma/2 - \mathrm{Im}\{\frac{V'+V_\mathrm{eff}}{2\hbar} \} )^2 } ,\\
    &\approx \frac{ 3 }{(\omega_1+\mathrm{Re}\{\frac{V'+V_\mathrm{eff}}{2\hbar}\} -\omega_\mathrm{D} )^2 + (\Gamma/2 - \mathrm{Im}\{\frac{V'+V_\mathrm{eff}}{2\hbar} \} )^2 } .
\end{align}

\end{appendix}

\newpage

\bibliography{references}

\end{document}